\documentclass[11pt,a4paper]{article}
\pdfoutput=1
\usepackage{epsfig}
\usepackage{latexsym}
\usepackage{amsfonts}
\usepackage{amsmath}
\usepackage{amsthm}
\usepackage{amssymb}
\usepackage{amsbsy}
\usepackage{multirow}
\usepackage{slashed}
\usepackage{color}
\usepackage{mathrsfs}
\usepackage{subfigure}
\usepackage[font={footnotesize,sl},bf]{caption}
\usepackage{wrapfig}

\usepackage{jheppub}

\newcommand{\bQ}{Q}
\newcommand{\pQ}{q}

\def\calb         {{\cal B}}

\def\cale         {{\cal E}}

\def\calh         {{\cal H}}

\def\call         {{\cal L}}

\newsavebox{\uuunit}
\sbox{\uuunit}
    {\setlength{\unitlength}{0.825em}
     \begin{picture}(0.6,0.7)
        \thinlines
        \put(0,0){\line(1,0){0.5}}
        \put(0.15,0){\line(0,1){0.7}}
        \put(0.35,0){\line(0,1){0.8}}
       \multiput(0.3,0.8)(-0.04,-0.02){12}{\rule{0.5pt}{0.5pt}}
     \end {picture}}

\def\be{\begin{equation}}
\def\ew{\end{equation}}
\def\bea{\begin{eqnarray}}
\def\eea{\end{eqnarray}}

\newcommand{\beq}{\begin{eqnarray}}
\newcommand{\eeq}{\end{eqnarray}}

\newcommand{\eal}[1]{\begin{equation} \begin{aligned} #1 \end{aligned}\end{equation}}

\def\a{\alpha}

\def\d{\delta}

\def\f{\phi}

\newcommand{\ti}[1]{\tilde #1}

\def\to{\rightarrow}

\def\sF{{{ F}\!\!\!\!\hskip.8pt\hbox{\raise1pt\hbox{/}}\,}}
\def\som{{{ \omega}\!\!\!\!\hskip.8pt\hbox{\raise1pt\hbox{/}}\,}}
\def\sJ{{{\rm J}\!\!\!\!\hskip.8pt\hbox{\raise1pt\hbox{/}}\,}}

\newcommand{\SO}{\mathop{\rm SO}}

\newcommand{\pd}[2]{\frac{\partial #1}{\partial #2}}
\newcommand{\fd}[2]{\frac{\delta #1}{\delta #2}}

\newcommand{\bdm}{\begin{displaymath}}
\newcommand{\edm}{\end{displaymath}}
\newcommand{\nn}{\nonumber \\}
\renewcommand{\f}[2]{\frac{#1}{#2}}

\title{Phases of non-extremal multi-centered bound states}
\author[a]{Borun D.\ Chowdhury,}
\author[b]{Daniel R.\ Mayerson,}
\author[c]{Bert Vercnocke}
\affiliation[a]{Department of Physics,\\
Arizona State University, Tempe, Arizona 85287, USA
}
\affiliation[b]{Institute for Theoretical Physics, University of Amsterdam,\\
Science Park 904, Postbus 94485, 1090 GL Amsterdam, The Netherlands}
\affiliation[c]{Institut de Physique Th\'eorique, CEA Saclay,\\  91191 Gif sur Yvette, France}

\emailAdd{borun.chowdhury@asu.edu}
\emailAdd{d.r.mayerson@uva.nl}
\emailAdd{bert.vercnocke@cea.fr}

\abstract{We investigate the phase space of multi-centered near-extremal configurations previously studied in arXiv:1108.5821 \cite{Anninos:2011vn} and arXiv:1110.5641 \cite{Chowdhury:2011qu} in the probe limit. We confirm that in general the energetically favored ground state of the multi-center potential, which can be a single or multi-center configuration, has the most entropy and is thus thermodynamically stable. However, we find the surprising result that for a subset of configurations, even though a single center black hole seems to be energetically favored, it is entropically not allowed (the resulting black hole would violate cosmic censorship). This disproves classical intuition that everything would just fall into the black hole if energetically favored. Along the way we highlight a shortcoming in the literature regarding the computation of the angular momentum coming from electromagnetic interaction in the probe limit and rectify it. We also demonstrate that static supertubes can exist inside ergoregions where ordinary point particles would be frame dragged.
}
\keywords{Black Holes in String Theory, AdS/CFT Correspondence, Black Holes}

\begin{document}
\maketitle

\section{Introduction}

Multi-centered BPS black holes \cite{Denef:2000nb,Bates:2003vx,Bena:2004de, Gutowski:2004yv,Gauntlett:2004qy,Elvang:2004ds,Bena:2005ni, Bena:2005va,
Berglund:2005vb,Saxena:2005uk} and extremal non-BPS multi-centered black holes  \cite{Goldstein:2008fq, Galli:2009bj,Bena:2009en,Bena:2009ev,Bena:2009fi,Bossard:2009my,Bossard:2009mz,DallAgata:2010dy,Bobev:2011kk,Bossard:2011kz,Bobev:2012af,Bossard:2012xsa,Bossard:2013oga} have been known in string theory for some time. These solutions are characterized by bubble equations which determine the relative locations of the centers. For these solutions, brane probes  capture  the same information as the fully backreacted solution. Taking one of the centers to be a probe supertube,\footnote{One typically considers supertubes, as such it is related by spectral flow to the most general smooth zero entropy charge combination~\cite{Bena:2008wt}.} one finds that the minimum of the supertube potential exactly reproduces the equilibrium distance set by the supergravity bubble equations, both for BPS solutions~\cite{Bena:2008dw} and for certain classes of extremal non-BPS solutions~\cite{Bena:2013gma}.

While a lot of physics of multi-centered solutions has been understood for extremal solutions, non-extremal solutions are much richer because they radiate and are thus more realistic. Furthermore, they may shed light on important issues like the information paradox~\cite{Hawking:1974sw} (see~\cite{Mathur:2009hf} for a recent review). Dealing with full blown non-extremality is expected to be a hard problem. Instead near-extremal solutions can be  studied for better control. By continuity, one would expect multi-centered solutions to exist for near-extremal systems also, albeit the minima of the potential would be lifted from the marginal value, the lifting governed by the amount of non-extremality. Indeed, such classically stable bound states were discovered in \cite{Anninos:2011vn,Bena:2011fc,Chowdhury:2011qu,Bena:2012zi,Chowdhury:2013pqa} with one or several centers fully backreacted and one center treated as a probe. This kind of analysis is helpful in the context of the fuzzball proposal as well, for the construction of non-extremal microstate geometries (see~\cite{Mathur:2005zp,Bena:2004de,Skenderis:2008qn,Balasubramanian:2008da,Chowdhury:2010ct} for a review of the fuzzball proposal). Only a handful of very specific backreacted non-extremal solutions are known at this time~\cite{Jejjala:2005yu,Giusto:2007tt,Bena:2009qv, Bobev:2009kn}; see \cite{Bena:2011fc,Bena:2012zi} for new probe constructions.

We will focus on non-extremal multi-centered bound states by putting probe supertubes in non-extremal black hole backgrounds. In \cite{Chowdhury:2011qu}, two of us proposed that rotating black holes will emit objects like supertubes to increase their entropy, quite like the Penrose process. It was further argued that the signature of this instability would be the minimum of the potential being lower than the value at the horizon. Such bound states were indeed found in \cite{Chowdhury:2011qu} (see also \cite{Anninos:2011vn}). We demonstrate representative potentials of this kind in Figure~\ref{fig:NearExtremal_Hamiltonian}.

While these findings based on the probe potential are suggestive of an interesting phase diagram, to say anything definite a careful analysis of the statistical weight of different configurations is required. This subtlety was discussed in~\cite{Anninos:2011vn}, but the analysis was done in the canonical ensemble. In applications where the black hole acts as a thermal bath, such that its temperature does not change during the process under consideration, the canonical ensemble is appropriate. However, when comparing stability of single center configurations towards forming multi-centered configurations and studying mergers of other centers with a black hole, the temperature of the black hole does change in general.

In this article we perform the analysis in the micro-canonical ensemble keeping energy, charges and angular momenta fixed during (de-)mergers. One subtlety that comes up when performing such an analysis is the determination of the angular momentum originating from the interaction between the electric charge of the background and the magnetic charge of the probe. The angular momentum of supertube probes in the supersymmetric BMPV black hole~\cite{Breckenridge:1996is} background was studied in the context of mergers in \cite{Marolf:2005cx,Bena:2008dw}, but the expressions used were incorrect as they did not transform covariantly under Lorentz transformations. We explain a correct procedure based on carefully applying the Noether method in the main body of this article, but we can already give the source of the error here. The electromagnetic angular momentum due to a magnetic monopole, $m$, and an electric point charge, $q$, located at $\vec \rho$ from the former can be easily found to be
\begin{equation}
\vec j = \int d^3 r~ \vec r \times (\vec E \times \vec B) = -\f{qm}{4 \pi} ~\vec \rho.
\end{equation}
However, if one naively tries to calculate the same in the probe limit from the static Lagrangian  $q \int dt A_i \dot x^i$ using the procedure for Noether charge, the z-component turns out to be
\begin{equation}
(j_z)_{naive} = q A_\phi = -\f{qm}{4 \pi}  (\cos \theta - k)\,,
\end{equation}
which is gauge dependent through the constant $k=\pm1$: the gauge field is ill-defined on the north or south pole of the sphere, depending on the choice of sign. Obviously, $(j_z)_{naive}$ cannot be correct. We show the covariant procedure in the probe limit in the next section and generalize this gauge-independent procedure to extended objects. It was an implicit gauge dependence which marred the computation of angular momentum in \cite{Marolf:2005cx,Bena:2008dw}.

We also find a curious feature that the ergoregion of a supertube, the region where it cannot remain static and is dragged along an angular direction, is different from the ergoregion for point-particles. Usually, the ergoregion is thought to be a property of the background and not of the background-and-probe system. This feature applies to four-dimensional probes as well. Wrapped brane probes become charged point particles with a position dependent mass, and hence the ergoregion for such a particle can depend on the embedding coordinates of the probe and not just on the background metric.

The main result of this paper builds on the carefully derived expression for the angular momentum. We use the angular momenta of probe branes to study the phase diagrams of multi-centered configurations graphically. In the micro-canonical ensemble a dominant phase is the one with more entropy. We find that dynamical stability implies thermodynamic stability (entropic dominance) but not vice versa. What was referred to as stable bound states in our earlier work \cite{Chowdhury:2011qu}, the ones with the potential at the minimum lower than that at the horizon (red curve in Figure~\ref{fig:NearExtremal_Hamiltonian_a}), are indeed stable in a thermodynamic sense. The story for the bound states at a local minimum of the probe potential that is higher than that at the horizon (green curve in Figure~\ref{fig:NearExtremal_Hamiltonian_b}) is not so straightforward. These were referred to as metastable in \cite{Chowdhury:2011qu} (in a quantum tunneling sense), but to be metastable the single center configuration formed by merging the supertube with the black hole should have more entropy. We however find that while this is the case in most of the region in phase space, there are some regions where this is not true. This means that in such regions, even though from the potential it seems that the supertube would want to tunnel through the barrier to fall into the black hole, the black hole is {\em not big enough}, in a phase space sense, to accommodate it. Since the entire phase space consists of not just one or two but multi-center configurations, it is not possible to scan over all of them to say what the end point would be, but it is certainly interesting to see that while the potential makes it seem like a merger is not only possible but likely, the story is very different. We also find regions where the potential would suggest the centers would merge, but a single center is simply not allowed for the corresponding charges. This is surprising because classically it seems a black hole would absorb everything. Conversely, there are regions in which stable two center configurations exist but no corresponding black hole which would have ``spat out" the supertube center.

Our results also give a glimpse of the  phase space of non-extremal black holes and black rings in Taub-NUT, as our two-center solutions can be related to non-extremal three-charge black rings in Taub-NUT by spectral flow \cite{Bena:2011zw}. We comment on this in the conclusion.

The plan of this paper is as follows. In Section~\ref{s:Gauge} we derive the angular momentum for an extended electric probe  moving in the background of a magnetic monopole. In Section~\ref{s:SupertubeProbe} we use the results of the preceding section to write down the potential and angular momentum of a supertube in a Cvetic-Youm black hole. In Section~\ref{s:Ergoregion} we demonstrate the curious feature that the ergoregion for a supertube differs from that of a point particle. In Section~\ref{s:PhaseSpacePlots} we plot the phase space for single and two-center configurations with fixed energy, charge and angular momentum. We conclude in Section~\ref{s:Conclusions}.

\section{Angular momenta of a probe}\label{s:Gauge}

In this section, we discuss the conserved angular momentum of a probe in a background with a magnetic field. Naively, the angular momentum depends on the background gauge potential, which is not gauge invariant. We discuss the procedure to find the correct gauge invariant conserved angular momentum.
For reasons of clarity, we explain the procedure in detail for a point particle in four dimensions (inspired by \cite{PortuguCs:2009fk}), and then generalize to a $p$-brane in arbitrary spacetime dimensions.

\subsection{Point particle in a magnetic field}

Consider a (non-relativistic) probe particle in the background of a magnetic monopole:
\begin{equation}
S = \int d\tau\left( \tfrac 12 M {\dot {\vec x}} ^2 + q A_i \dot x^i\right)\,,\label{eq:PP_Action}
\end{equation}
with the background magnetic potential
\begin{equation}
A = \f{m}{4 \pi} (k-\cos \theta) d \phi\,,
\end{equation}
with $m$ the magnetic monopole charge and $k=\pm1$, depending on the gauge choice. For instance $k=1$ gives a potential that is well defined on the north pole of the $S^2$ spanned by $(\theta,\phi)$, and there is a Dirac string on the negative $z$-axis $(\theta = \pi)$.

Since the background magnetic field is spherically symmetric, one would expect angular momentum to be conserved. However, with the conjugate momenta $p_i = \partial \call / \partial \dot x^i$, the ``naive angular momentum" around the $z$-axis is given by:
\begin{equation}
(j_z)_{naive} = p_\phi = M r^2 \sin^2\theta \dot \phi +q A_\phi\,.\label{eq:PP_AngMom}
\end{equation}
This is not covariant under rotations. Take for example a rotation around the $x$-axis, such that $\theta' =\pi - \theta$ and $\phi' = 2\pi - \phi$. Then the gauge potential becomes
\begin{equation}
A' =  \f{m}{4 \pi} (-\cos \theta - k) d \phi = A (k\to -k)\,.
\end{equation}
This rotates the position of the Dirac string. Of course, this is a gauge artifact and we conclude that one needs to supplement a rotation by a compensating gauge transformation to ensure that the angular momentum transforms covariantly under rotations. There are several ways to find the form of the compensating gauge transformation. One can for instance demand that the angular momentum transforms as a vector under rotations \cite{Coleman:1982cx}, or one can use the angular momentum form $\vec L = M \vec r \times \dot \vec r -\f{q m}{4 \pi} \vec r/r$ \cite{Wu:1976ge}. Another way, which we will use because of its straightforward extension to higher-dimensional probes, is to implement the gauge transformation in the Noether procedure for the construction of the angular momentum.

\subsubsection{Angular momentum from Noether procedure}

Under a general symmetry transformation of the embedding coordinates of the particle $x' = x + \delta x$, the variation of the Lagrangian must be a total derivative:
\begin{equation}
\delta L =\frac{\partial L}{\partial x^i} \delta x^i  +\frac{\partial L}{\partial \dot x^i} \frac{d \delta x^i}{d\tau}\equiv \frac{d K}{d \tau}\,.
\end{equation}
Using the Euler-Lagrange equations, this gives the conserved charge $dQ/d\tau=0$:
\begin{equation}
Q = p_i \delta x^i - K\,.
\end{equation}
In most applications, the contribution $K$ for the conserved charges associated to rotations is exactly zero and we find the standard expressions for the conserved (angular) momenta, as in eq.\ \eqref{eq:PP_AngMom}. For an electric particle in a magnetic field however, the contribution $K$ is  exactly the necessary compensating gauge transformation of the gauge field discussed earlier.

Take a rotation with infinitesimal generator $\delta x_I = \xi_I$, where the subscript $I$ labels the rotation axis. The first term in the Lagrangian \eqref{eq:PP_Action} is rotationally invariant. The second term gives the contribution:
\begin{equation}
\delta_{\xi_I} L = q(\call_{\xi_I} A_i )\,\dot x^i\equiv\frac{d K_I}{d \tau} \,,
\end{equation}
where $\call_\xi$ is the Lie derivative. To see that this equation really gives the total derivative of a function $K_I$, remember that the potential $A$ is gauge dependent. Hence it must only be left invariant by rotations up to a gauge transformation:
\begin{equation}
\delta_{\xi_I} A_i = \call_{\xi_I} A_i \equiv \partial_i \Lambda_I\,.
\end{equation}
Therefore we  find
\begin{equation}
K_I = q\Lambda_I\,,
\end{equation}
and the conserved angular momentum charge is
\begin{equation}
j_I = \xi^i_Ip_i  - q \Lambda_I \label{eq:ConservedAngMom}\,.
\end{equation}

Explicitly, the generators for rotations along the three axes are:
\begin{eqnarray}
\xi_X &=& - \sin \phi \frac{\partial}{\partial \theta}- \cos \phi \cot \theta \pd{}\phi\,,\nonumber\\
\xi_Y &=& \cos \phi \pd{}\theta - \sin \phi \cot \theta \pd{}\phi\,,\nonumber\\
\xi_Z &=& \pd {}\phi\,,
\end{eqnarray}
they satisfy $[\xi_X,\xi_Y]=-\xi_Z$ and cyclic permutations. The associated gauge transformations are
\begin{eqnarray}
\Lambda_X &=&-\f{m}{4 \pi}(k\cos\theta-1)\frac{\cos \phi}{\sin \theta}\,,\nonumber\\
\Lambda_Y &=&-\f{m}{4 \pi}(k\cos\theta-1)\frac{\sin \phi}{\sin \theta}\,,\nonumber\\
\Lambda_Z &=&\f{k m}{4 \pi}\,.
\end{eqnarray}
Note that the condition $\call_\xi A_i= \partial_i \Lambda_I$ does not fix the constants in the gauge transformations $\Lambda_I$ (in particular $\Lambda_Z = k$). We need to impose the Poisson brackets for the $\SO (3)$ algebra of rotations:
\begin{equation}
\{j_X,j_Y\} = - j_Z
\end{equation}
and cyclic in $X,Y,Z$. In particular, this gives the condition
\begin{equation}
\xi^i_X \partial_i \Lambda_Y-\xi^i_Y \partial_i \Lambda_X=-\Lambda_Z\,,
\end{equation}
which determines $\Lambda_Z$ completely. The final expression for the covariant angular momentum is then
\begin{equation}
j_z = Mr^2 \sin^2 \theta \dot \phi-\f{q m}{4 \pi} \cos \theta.
\end{equation}
As a check, the static part of this expression is also obtained by integrating $(\vec r \times (\vec E \times \vec B))_z$ over all space.

\subsection{Extended object in a magnetic field}

We can readily extend the discussion to $p$-branes in arbitrary spacetime dimensions. The action for a probe $p$-brane with charge $q$, in a background with a $p$-form magnetic potential is:
\begin{equation} \label{eq:extendedobj_action}
S= \int d^{p+1}\sigma \call \equiv\int  d^{p+1}\sigma \call_0 + q\int C_{p+1}\,,
\end{equation}
where $\sigma^\a, \a = 0,\ldots, p$ are the worldvolume coordinates, $\call_0$ denotes the other terms in the worldvolume action (we do not need their exact form for the present discussion) and the integral over the $(p+1)$-form is over the pullback on the $p$-brane's worldvolume:
\begin{equation}
\int C_{p+1} = \int d^{p+1}\sigma\,\frac 1 {(p+1)!} C_{i_1 \ldots i_{p+1}} \epsilon^{\alpha_1 \ldots \alpha_{p+1}}(\partial_{\a_1} x^{i_1})\ldots(\partial_{\a_{p+1}} x^{i_{p+1}})\,.
\end{equation}

Under an infinitesimal symmetry transformation of the embedding coordinates $x'(\sigma) = x(\sigma) + \delta x(\sigma)$, the Lagrangian must be invariant up to a total derivative. This gives:
\begin{equation}
\delta \call = \left(\frac{\partial \call}{\partial (\partial_\a x^i)} \frac{\partial \delta x^i}{\partial\sigma^\a} + \frac{\partial \call}{\partial x^i} \delta x^i\right) \equiv \partial_\a K^\a\,.
\end{equation}
Using the Euler-Lagrange equations, this gives the conserved current (with $p^{\alpha}_i\equiv \partial\mathcal{L}/\partial(\partial_{\alpha}x^i)$) :
\begin{equation}
j^\alpha = p^{\alpha}_i \delta x^i - K^\alpha\,,
\end{equation}
and the conserved charge (with $d^p\sigma\equiv d\sigma^1\ldots d\sigma^p$):
\begin{equation}
Q = \int d^{p}\sigma j^0 = \int d^{p}\sigma(p^0_i \delta x^i - K^0)\,.
\end{equation}
We could rewrite this in a reparametrization-invariant form, but for simplicity we will just assume that $\sigma^0$ is the timelike direction on the worldvolume so that we can integrate charges over surfaces of constant $\sigma^0$.

\subsubsection{Angular momentum from Noether procedure}
Consider the conserved charges for infinitesimal symmetry generators $\delta x = \xi_I$, labeled by $I$. We assume that the term $\call_0$ is invariant under the symmetry, such that only the gauge potential term transforms:
\begin{equation}
\delta_I \call = \frac q {(p+1)!} (\call_{\xi_I}C_{i_1 \ldots i_{p+1}}) \epsilon^{\alpha_1 \ldots \alpha_{p+1}}(\partial_{\a_1} x^{i_1})\ldots(\partial_{\a_{p+1}} x^{i_{p+1}}) \equiv \partial_\a K^\a_I\,.
\end{equation}
Just as for the point particle, the symmetry generators leave the gauge field invariant up to a gauge transformation:
\begin{equation}
\call_{\xi_I} C_{p+1} \equiv  d\Lambda^I\,,\label{eq:GaugeTransfo_PForm}
\end{equation}
where $\Lambda^I$ are $p$-forms of gauge transformations. We get
\begin{equation}
K^\a_I =  \frac q {p!}\Lambda^I_{i_1 \ldots i_p} \epsilon^{\a \alpha_1 \ldots \alpha_p}(\partial_{\a_1} x^{i_1})\ldots(\partial_{\a_{p}} x^{i_{p}}) \,.
\end{equation}
With $\epsilon^{012\ldots p} = -1$, this gives the conserved charges:
\begin{equation}
Q_I = \int d^{p}\sigma( p_i^0 \xi^i_I) + q\int \Lambda^I\label{eq:Qasintegral}\,.
\end{equation}
where the second term denotes the integral of the pull-back of $\Lambda$ on the same $\sigma^0 = cst$ surface as for the first integral.

As for the point particle, closed terms in the gauge transformations (terms for which $d \Lambda^I =0$) cannot be determined from (\ref{eq:GaugeTransfo_PForm}). They can be fixed by demanding that the Poisson brackets of the conserved charges satisfy the same symmetry algebra as the Lie brackets of the symmetry generators $\xi_A$:
\begin{equation}
[\xi_A,\xi_B] = f_{AB}{}^C \xi_C\,,\qquad \{Q_A,Q_B\} = f_{AB}{}^C Q_C\,.\label{eq:Poisson_Charges}
\end{equation}

The non-trivial components of the Poisson brackets of the conserved charges are
\footnote{Note that the Poisson brackets involve functional derivatives. For any two functionals $F = \int d^p\sigma\,f[\vec x(\sigma),\vec p(\sigma),\partial_\a \vec x(\sigma)]$ and $G = \int d^p\sigma\,g[\vec x(\sigma),\vec p(\sigma),\partial_\a \vec x(\sigma)]$\,, the Poisson brackets are
\begin{equation}
\{F ,G\} = \int d^p\sigma \left(\fd F {p_i(\sigma)} \fd G {x^i(\sigma)}-\fd F {x^i(\sigma)} \fd G {p_i(\sigma)}\right)\,.
\end{equation}
with
\begin{equation}
\fd F {x^i(\sigma)} = \pd f {x^i} - \pd {}{\sigma^\a} \left(\pd f {\partial_\alpha x^i}\right)\,, \qquad \fd F {p_i(\sigma)} = \pd f {p_i} \,, \label{tempLabel1}
\end{equation}
and analogously for $G$.
}
\begin{eqnarray}
\nonumber \{Q_A,Q_B\} &=& \int d^p\sigma (\xi_A^i\partial_i \xi_B^j - \xi_B^i\partial_i \xi^A_j)p^0_i\\
&  +& \int d^p\sigma
(p+1)!\, (\xi_A^i \partial_{[i} \Lambda^B_{i_1 \ldots i_p]}-\xi_B^i \partial_{[i} \Lambda^A_{i_1 \ldots i_p]}) \left(\epsilon^{ \alpha_1\ldots \alpha_p}\partial_{\alpha_{1}} x^{i_1}\ldots \partial_{\alpha_{p}} x^{i_p}\right)\,.
\end{eqnarray}
Since the first term equals $f_{AB}{}^C \xi^i_Cp_i^0$, the Poisson bracket equations \eqref{eq:Poisson_Charges} give the following constraint on the gauge parameters:
\begin{equation}
i_{\xi_A}(d\Lambda_B) -i_{\xi_B}(d\Lambda_A) = f_{AB}{}^C(\Lambda_C+d\lambda_C) \label{eq:Lambda_Gauge_orig}\,.
\end{equation}
We have allowed for an arbitrary $(p-1)$-form $\lambda_C$ on the right-hand side, since the gauge transformations $\Lambda$ are $p$-forms that have a ``gauge invariance'' themselves: $\Lambda_C \to \Lambda_C + d \lambda_C$; the term proportional to $d\lambda_C$ is a total derivative and will thus not contribute to the integral $Q_C$ as given in (\ref{eq:Qasintegral}).

\subsubsection{A String in five dimensions}
Let us work this out for an example. Consider a string in five-dimensional Minkowski spacetime, with spatial coordinates:
\begin{eqnarray}
x_1 = \sin \theta \cos \phi\,,&\qquad & x_3 = \cos \theta \cos \psi\,,\nonumber\\
x_2 = \sin \theta \sin \phi\,,&\qquad & x_4 = \cos \theta \sin \psi\,,
\end{eqnarray}
and a background magnetic field
\begin{equation}
C_2 = m(k-\cos^2\theta) d\phi\wedge d\psi\,.
\end{equation}
We choose worldvolume coordinates $\sigma^0 = \tau, \sigma^1 = \sigma$.

We concentrate on the conserved charges for rotations in the 12 and 34 planes. From the Noether procedure, we find these are:
\begin{equation}
Q_{12} = \int d\sigma(p^{\tau}_\psi + \Lambda^{12}) \,,\qquad Q_{34} = \int d\sigma(p^{\tau}_\psi + \Lambda^{34})\,.
\end{equation}
with  $d\Lambda^{12} =  d\Lambda^{34} = 0$. By demanding that all of the angular momentum charges obey the $SO(4)$ algebra (see Appendix \ref{app:SO(4)} for more details),
\begin{equation}
\{Q_{ik},Q_{j\ell}\} = \delta_{i\ell}Q_{kj}- \delta_{k\ell}Q_{ij}+ \delta_{kj}Q_{i\ell}- \delta_{ij}Q_{k\ell}\,,
\end{equation}
we find the one-forms:
\begin{equation}
\Lambda^{12}=m k\, d\psi\,,\qquad \Lambda^{34} = m(1-k)\,d\phi\,.
\end{equation}
Note that even though these one-forms satisfy are closed, $d\Lambda^{13} = d \Lambda^{34} =0$, they are not globally exact and thus not pure gauge: there is no globally well-defined $(p-1)$ form $\lambda$ which can transform them to zero as $\Lambda\rightarrow \Lambda+d\lambda$.

The gauge-independent conserved charges are then:
\begin{equation}
Q_{12} = Q_{12}^0+\int d\sigma\left(m\cos^2 \theta\right)\partial_{\sigma}\psi \,,\qquad Q_{34} = Q_{34}^0+\int d\sigma\left(m\sin^2 \theta\right)\partial_{\sigma}\phi
\end{equation}
Here $Q_{ij}^0$ denotes the orbital angular momentum (the part coming from $\call_0$ in \eqref{eq:extendedobj_action}).

We will make use of this result in the following section, where we consider supertubes in a non-extremal black hole background with a background magnetic field.

\section{Supertube probe in a non-extremal black hole background} \label{s:SupertubeProbe}

In this section, we give the potential and angular momenta for a supertube in the background of the five-dimensional Cvetic-Youm black hole. We use the discussion of the previous section to obtain the gauge invariant angular momenta.

\subsection{Background}

The Cvetic-Youm black hole \cite{Cvetic:1996xz,Cvetic:1996kv,Giusto:2004id} is a non-extremal, rotating three charge black hole of five-dimensional supergravity. It has two angular momenta in two independent planes in $\mathbb{R}^4$. We give the solution in the M-theory frame where it arises from a $T^6$ compactification. The three charges come from M2 branes wrapped on three orthogonal $T^2$'s inside $T^6$.

The solution depends on six parameters: $m$ encodes the temperature, the three `boosts' $\delta_I$ control the charges and $a_1,a_2$ determines the angular momenta. The metric and gauge field are
\begin{eqnarray}
ds_{11}^2 &=& -(H_1 H_2 H_3)^{-2/3} H_m (dt + k)^2 + (H_1 H_2 H_3)^{1/3} ds_4^2 + \sum_{I=1}^3 \frac{(H_1 H_2 H_3)^{1/3}} {H_I}ds_I^2\,,\nn
A_3 &=& \sum_{I=1}^3 A^{(I)} \wedge \omega_I\,,\qquad A^{(I)} = \coth (\delta_I) H_I^{-1} (dt + k) + B^{(I)} - \coth (\delta_I) dt\,,\label{eq:11d_Background}
\end{eqnarray}
where $ds_I^2$ and $\omega_I$ are the flat metric and volume form on the $I^{\rm th}$ torus. The rotation one-form $k$ and magnetic parts $B^{(I)}$ of the gauge fields are\footnote{Note that $B^{(I)}$ blows up in the zero charge limit $\delta_I\to 0$. This is an artefact of the form we chose to present the gauge fields in \eqref{eq:11d_Background}. The fields $B^{(I)}$ appear through the actual physical gauge fields $A^{(I)}$, which do vanish in this limit (the $B^{(I)}$-term cancels the divergent contribution coming from $k$ in eq.\ \eqref{eq:11d_Background}).}
\begin{eqnarray}
\label{eq:kvector} k &=& \f{m}{f} \left[ -\frac{ c_1 c_2 c_3}{H_m} (a_1 \cos^2\theta\, d\psi + a_2\sin^2 \theta\, d\phi) + s_1s_2s_3 (a_2 \cos^2\theta\, d\psi + a_1\sin^2 \theta \,d\phi) \right]\,\nonumber,\\
B^{(I)}&=& \frac{m}{f H_m} \frac{ c_J c_K}{s_I} (a_1 \cos^2\theta \,d\psi + a_2\sin^2 \theta d\phi)\,,
\end{eqnarray}
with $I,J,K$ all different and we write
\begin{equation}
c_I \equiv \cosh \delta_I\,, \qquad s_I \equiv \sinh \delta_I\,.
\end{equation}

The four-dimensional base metric is
\begin{eqnarray}
ds_4^2 &=& \frac{f r^2}{g}dr^2 + f ( d \theta^2 + \sin^2\theta\, d \phi^2 + \cos^2 \theta\, d\psi^2)\nn
&&+ H_m^{-1} (a_1 \cos^2\theta \,d\psi + a_2\sin^2 \theta d\phi)^2 - (a_2 \cos^2\theta \,d\psi + a_1\sin^2 \theta \,d\phi)^2\,.
\label{eq:4d_Base}
\end{eqnarray}
The solution is built from the functions
\begin{eqnarray}
&&H_I = 1 + \frac{m s_I^2}{f}\,,\qquad H_m = 1 - \frac{m}f\,,\qquad f = r^2 + a_1^2 \sin^2\theta + a_2^2 \cos^2 \theta\,,\nn
&&g = (r^2 + a_1^2)(r^2 + a_2^2) - m r^2 \equiv (r^2 - r^2_+)(r^2 - r^2_-)\,.
\end{eqnarray}
The roots of the function $g(r)$ give the radial position of the inner and outer horizon:
\begin{equation}
(r_\pm)^2 = \frac 12 \left( m- {a_1^2}-{a_2^2} \pm \sqrt{\left(m- a_1^2-a_2^2\right)^2-4 a_1^2 a_2^2}\right)\,.
\end{equation}
The ADM mass, electric charges and angular momenta of the black hole are (in units where $G_5=\pi/4$):
\eal{
M_{ADM} &= \frac m 2 \sum_I \cosh 2 \delta_I\,, \qquad &J_\psi =-m( a_1 c_1 c_2 c_3 - a_2 s_1 s_2 s_3)\,,\\
\bQ_I &= \frac m 2 \sinh 2 \delta_I\,, &J_\phi =-m( a_2 c_1 c_2 c_3 - a_1 s_1 s_2 s_3)\,.
\label{eq:5dCharges}
}
There are two extremal limits. The supersymmetric extremal limit is $m, a_1,a_2 \to 0$ and $|\delta_I| \to  \infty $  while keeping fixed the charges $\bQ_I$ and ratios $a_i/\sqrt{m}$. The four-dimensional base space becomes flat and one recovers the supersymmetric rotating three-charge BMPV black hole \cite{Breckenridge:1996is} with $M_{ADM} = \sum_I |\bQ_I|$. In the rest of this paper, we reserve the term ``supersymmetric limit'' for the choice $Q_I>0$. The non-supersymmetric extremal limit is obtained by putting $m = (|a_1| +|a_2|)^2$ and has $M_{ADM} > \sum_I |\bQ_I|$. This is the `ergo-cold' black hole studied in \cite{Dias:2007nj}.

\subsection{Potential and angular momentum of a supertube}
We consider supertubes with the two charges $\pQ_1$ and $\pQ_2$ corresponding to M2 branes on the first two $T^2$'s. We use lower case for probe charges, upper case for background charges.  The dipole charge, which we call $d_3$, is an M5 brane along those two $T^2$'s and along a one-cycle in the four-dimensional base which we parameterize by an angular coordinate $\a$  and two constants $b_1,b_2$ describing its embedding as
\begin{equation}
\psi = b_1 \a\,, \quad\phi = b_2 \a\,.
\end{equation}

The supertube potential is (see appendix \ref{app:AngularMomentum} and \cite{Chowdhury:2011qu}):
\begin{eqnarray}
\calh &=&\f{1}{|d_3|} \frac{\sqrt{H_m H_1H_2H_3 g^{(4)}_{\a \a}}}{R^2}\sqrt{\ti \pQ_1^2 + d_3^2 \frac{R^2}{H_2^2}}\sqrt{\ti \pQ_2^2 + d_3^2 \frac{R^2}{H_1^2}}+\f{1}{d_3} \frac{H_m k_\a}{R^2} \ti \pQ_1 \ti \pQ_2 - d_3 m  \f {\bQ_3}{\bQ_1 \bQ_2} B^{(3)}_\a\nonumber\\
&&-\coth \d _1\left(\f{ \tilde \pQ_1}{H_1}-\pQ_1\right)    - \coth \d _2 \left(\f{ \tilde \pQ_2}{ H_2}-\pQ_2\right)- d_3 \coth \d _1 \coth \d _2 \frac{k_\a}{H_1 H_2}\,,
\label{eq:NonExtremalHamFullBulk}
\end{eqnarray}
where $k_\a,g^{(4)}_{\a \a},B^{(3)}_\a$ are the pullbacks of the rotation one-form, the four-dimensional metric (\ref{eq:4d_Base}) and the third magnetic field on the supertube worldvolume. The two kinds of charges appearing above are related as
\begin{equation}
\ti \pQ_1 = \pQ_1 - d_3 A_\a^{(2)}\,,\qquad \ti \pQ_2 = \pQ_2 - d_3 A_\a^{(1)}\,, \label{ShiftedCharges}
\end{equation}
where $A^{(I)}_\alpha$ are the pullbacks of the gauge fields on the supertube worldvolume. Note  that  $\ti \pQ_1$ and $\ti \pQ_2$ are the brane source charges which are not conserved or quantized but $q_1$ and $q_2$ are the Page  charges which are conserved and quantized (see \cite{Page:1984qv,Chowdhury:2013pqa, deBoer:2012ma, Marolf:2000cb}). Thus the latter quantities will have to be kept track of when discussing supertube and black hole mergers. We have also introduced the square radius:
\begin{equation}
R^2 \equiv H_1H_2H_3 g^{(4)}_{\a \a} - H_m k_\a^2\,.
\end{equation}

The angular momenta of the supertube are (see appendix \ref{app:AngularMomentum})
\begin{eqnarray}
\frac{j_i}{T_{D4}} &=&  \frac 1 {d_3}\frac{g_{\a i}^{(4)}k_\a-k_i g_{\a\a}^{(4)}}{g_{\a\a}^{(4)} }\frac{\sqrt{H_1H_2H_3H_mg_{\a\a}^{(4)}}}{R^2}\sqrt{\tilde \pQ_1^2 +d_3^2 \frac{R^2}{H_2^2}}\sqrt{\tilde \pQ_2^2 +d_3^2 \frac{R^2}{H_1^2}} \nonumber\\
&&+\frac{g_{\a i}^{(4)}Z^3- k_i k_\a H_m}{R^2} \frac{\tilde \pQ_1 \tilde \pQ_2}{d_3}
 +d_3 b_j (-C_{ij} + \kappa_i) +\tilde \pQ_2 A_i^{(2)} + \pQ_1 A_i^{(1)} \,,\label{eq:AngularMomentum}
\end{eqnarray}
where $i,j$ run over $\psi$, $\phi$; $k_i$ are the (non-pulled back) components of $k$ in \eqref{eq:kvector}, $g^{(4)}$ again stands for the 4D metric (\ref{eq:4d_Base}) and the two-form components appearing in this expression are
\begin{equation}
C_{\psi\phi} = -\f{Q_3}{f H_2}\left[ (r^2 + a_2^2 + m s_2^2) \cos^2 \theta - \f{M(s_2^2-s_1^2)}{fH_1}(a_1^2-a_2^2)\cos^2 \theta \sin^2 \theta\right]\,,
\end{equation}
We have defined the constants
\begin{equation}\label{eq:angmomkappas}
\kappa_\psi = -Q_3\,,\qquad \kappa_\phi = 0\,.
\end{equation}
We derive the form of the angular momentum from a DBI treatment in Appendix \ref{app:AngularMomentum}.  The constants $\kappa_i$ are determined by demanding that the angular momentum charges in the flat space limit, or equivalently at spatial infinity, satisfy the $SO(4)$ algebra as discussed in Section \ref{s:Gauge}. A non-trivial check of the constants $\kappa_i$ fixing the gauge ambiguity, is that the angular momentum is symmetric under the unphysical relabeling $(b_1,\psi,\theta) \leftrightarrow (b_2,\phi,\pi/2-\theta)$.

At first sight, the angular momenta do not seem to be symmetric under interchange of the tori 1 and 2 (while the Hamiltonian clearly is). However, a closer look shows that this symmetry of the supertube physics is present: the antisymmetric terms, residing solely in terms in the last line of \eqref{eq:AngularMomentum}, nicely cancel when expanding those terms.

\subsection{Comparison with the literature}

We can rearrange the angular momenta in its physically interesting components: the part along the supertube $j_{\|}$ and the part transverse to its worldvolume $j_{\perp}$:
\begin{equation}
j_{\|} \equiv b_1 j_\psi + b_2 j_\phi \,,\qquad j_\perp \equiv b_1 j_\phi - b_2 j_\psi \,.
\end{equation}
The parallel component takes the particularly simple form
\begin{equation}
j_\| = \frac{q_1q_2}{d_3} -b_1 b_2 d_3 Q_3\,.\label{eq:AngMom_Parallel}
\end{equation}
The transverse angular momentum is not very elucidating. We only explicitly give two interesting limits. For the probe embedding we will use later, $\theta =0$ and $b_2=0$, only the gauge field term has a non-zero contribution, irrespective of the supertube position:
\begin{equation}
j_\perp^{(b_2=\theta=0)} = b_1 j_\phi = b_1d_3 Q_3\,.\label{eq:AngMom_perp_1}
\end{equation}

Second, we discuss the full expression for a supersymmetric background (BMPV black hole), evaluated at the supersymmetric bound state:
\begin{equation}
j_\perp^{\rm susy} = -d_3 Q_3 (b_1^2\cos^2 \theta - b_2^2\sin^2 \theta) - \frac{q_1q_2}{d_3} b_1 b_2(\cos^2 \theta - \sin ^2\theta)\,.
\end{equation}
The angular momenta of a supersymmetric black hole-supertube bound state have appeared before. However, only the full backreacted solution gives the correct result. See for instance \cite{Bena:2011zw} for a detailed account on the asymptotic charges of the two-center bound state. In our conventions, these are supertubes in the BMPV background with embedding $b_1=-b_2=1$, corresponding to a supertube along the Gibbons-Hawking fibre.
% \footnote{To compare to the earlier account of \cite{Bena:2005zy}, eq.\ (5.7) take the embedding $b_1=1,b_2=0$ and $\alpha = \tan \theta$.}
The angular momenta are:
\begin{equation}
j_{\|}^{\rm susy} = d_3 Q_3 + \frac{q_1 q_2}{d_3}\,,\qquad j_{\perp}^{\rm susy} = \left(-d_3 Q_3 + \frac{q_1 q_2}{d_3}\right) \cos(2 \theta) \,.
\end{equation}
Note that $j_\perp^{GH}$ is the symplectic product of the charge vectors of the black hole and the supertube.
For probe supertubes in supersymmetric black hole backgrounds, $j_\perp^{\rm susy}$ has been computed in \cite{Bena:2008dw} and \cite{Marolf:2005cx} without fixing the gauge ambiguity discussed in Section~\ref{s:Gauge}.\footnote{The authors of \cite{Bena:2008dw} noted that the gauge-dependent charges computed from the Noether procedure do match the charges in the harmonic functions (termed ``Gibbons-Hawking charges'') when the black hole and the supertube coalign on the three-dimensional base of Taub-NUT.}

\section{Ergoregions for supertubes different from those of point particles} \label{s:Ergoregion}

In this section we demonstrate explicitly a curious feature -- that the existence of an ergoregion is not just a background property, but can depend on the details of a probe in the background as well. This would make it possible for a supertube to be static inside the region where a point particle cannot be. To this end, we compare the ergoregion for probe particles to that of probe supertubes in the Cvetic-Youm background.

The black hole ergoregion is defined as the region of spacetime outside the horizon where every asymptotically timelike Killing vector becomes spacelike. For the Cvetic-Youm black hole in the M-theory frame, the relevant asymptotically timelike Killing vector is $\partial/\partial t$ and the ergoregion is\footnote{In principle there is a continuous family of such Killing vectors of the form $\partial_t + v^i \partial_i$, where $i$ runs over the compact directions and $|v^i| <1$. By symmetry the minimum region will be for $v^i=0$. As explained in \cite{Jejjala:2005yu,Cardoso:2007ws}, when there is broken symmetry by having momentum along one of the torus directions, the correct procedure is to boost to a frame where the momentum become zero to get the ergoregion.}
\begin{equation}
r_+ < r < r_{\rm erg}(\theta)\,,\label{eq:ergoregion_CY}
\end{equation}
with $r_+$ the outer horizon radius and $H_m(r_{\rm erg}(\theta))=0$:
\begin{eqnarray}
(r_+)^2 &=& \frac 12 \left( m- {a_1^2}-{a_2^2} + \sqrt{\left(m- a_1^2-a_2^2\right)^2-4 a_1^2 a_2^2}\right)\,,\nn
(r_{\rm erg})^2 &=& m -  a_1^2 \sin^2\theta - a_2^2 \cos^2 \theta\,.
\end{eqnarray}
In the ergoregion, a point particle cannot be held static.  If we were to insist on a static worldline, the particle's action would be complex. The wordline action of a static point particle in a gravitational background is:
\begin{equation}
S_{pp} = -M \int dt \sqrt{-g_{tt}}\,.
\end{equation}
so that the ergoregion is defined to be the region where $g_{tt}>0$. For a point particle in the Cvetic-Youm metric \eqref{eq:11d_Background}, this agrees with the  ergoregion defined as in \eqref{eq:ergoregion_CY}.

\begin{figure}[hbt!]
\centering
\subfigure[$b_1=1,b_2=0$]{
\includegraphics[width=.25\textwidth]{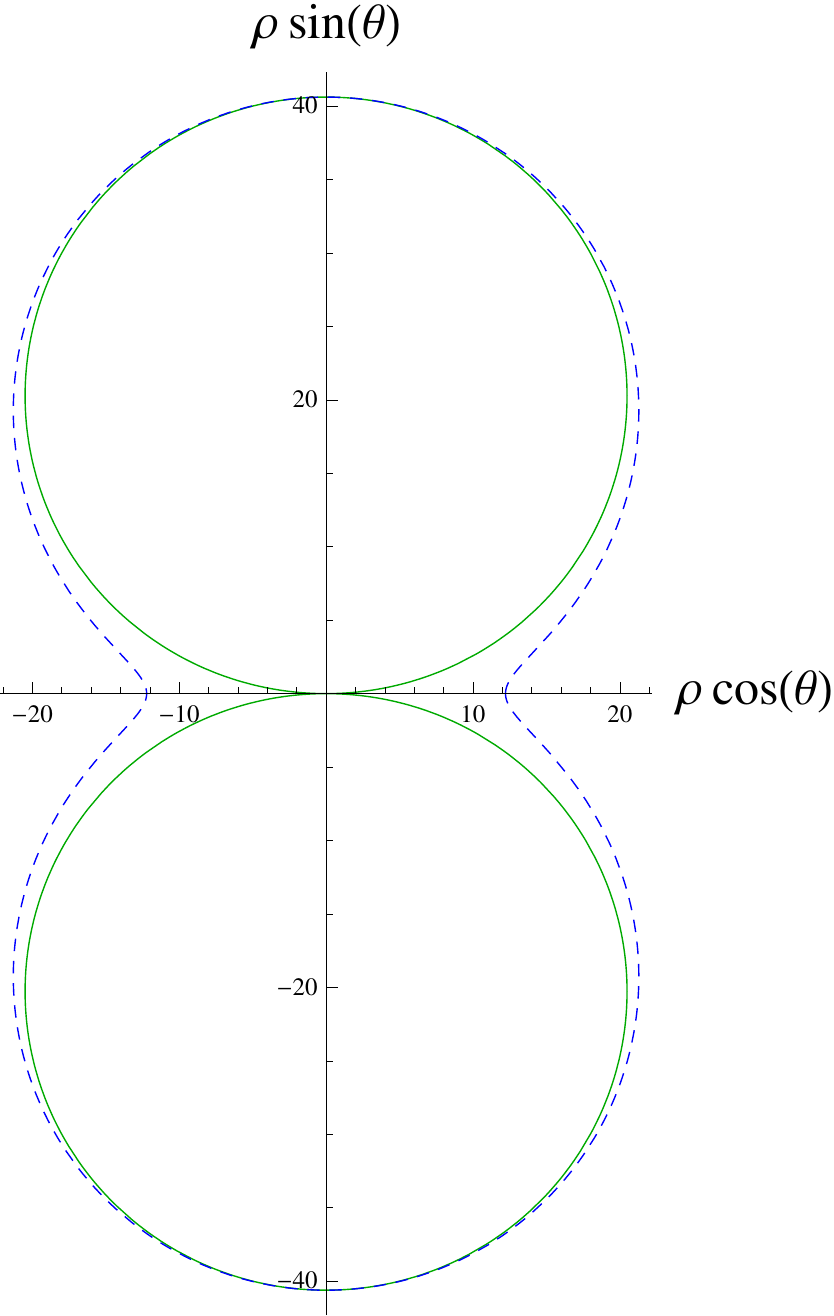}
}
\hspace{.05\textwidth}
\subfigure[$b_1=0,b_2=1$]{
\includegraphics[width=.25\textwidth]{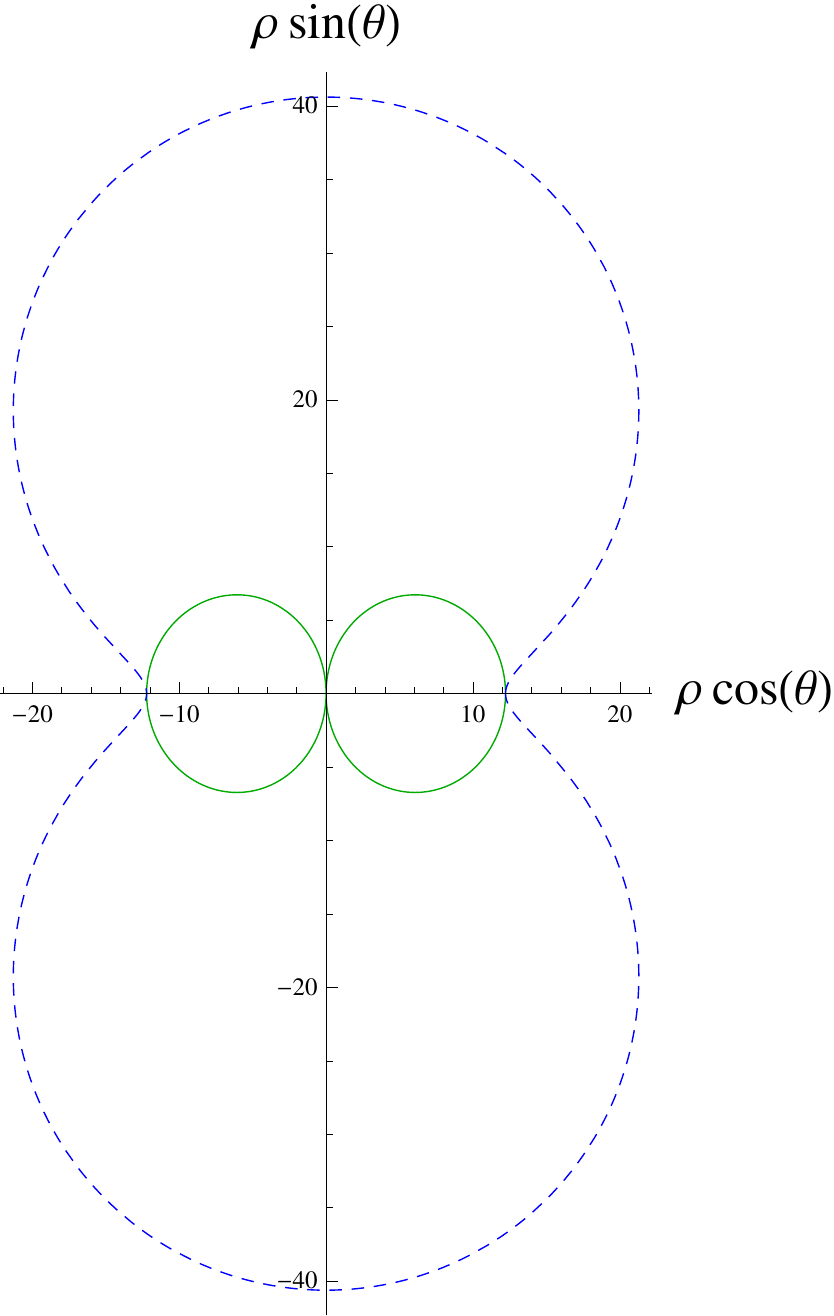}
}
\hspace{.05\textwidth}
\subfigure[$b_1=1,b_2=1$]{
\includegraphics[width=.25\textwidth]{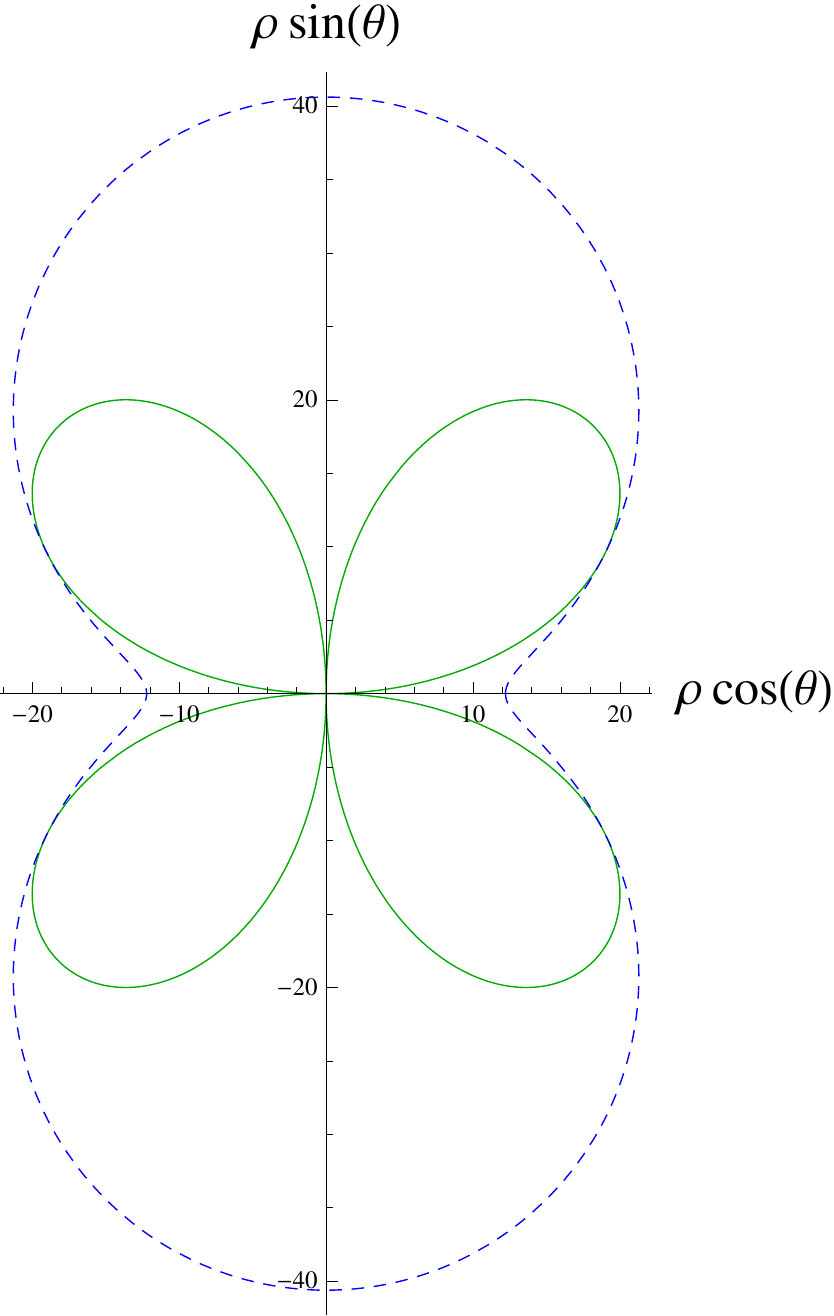}
}
\caption{\small Comparison of the ergoregions for point particles (blue, dashed) and supertubes (green, full) for different supertube embeddings. The background parameters used in the plots are $m=5000,a_1=10,a_2=-40$. We give the ergoregions in the $(\rho,\theta)$-plane, with $\rho = \sqrt{r^2-r^2_+}$ (horizon is at $\rho = 0$).\label{fig:Ergoregions}}
\end{figure}

The supertube is an extended object; its wordvolume potential involves more metric components. The ``supertube-ergoregion", which we define as the region outside the horizon where the potential for a static supertube is not defined, can depend on the way the supertube is embedded in spacetime. The relevant term of the supertube potential \eqref{eq:NonExtremalHamFullBulk} is:
\begin{equation}
\f{1}{|d_3|} \frac{\sqrt{H_m H_1H_2H_3 g^{(4)}_{\a \a}}}{R^2}\sqrt{\ti \pQ_1^2 + d_3^2 \frac{R^2}{H_2^2}}\sqrt{\ti \pQ_2^2 + d_3^2 \frac{R^2}{H_1^2}}\,.
\end{equation}
The necessary condition for the potential to be real is
\begin{equation}
H_m g^{(4)}_{\a \a} \geq 0\,.
\end{equation}
This condition is the analog of  $H_m>0$ for a point particle.\footnote{One could object that in principle $R^2 = H_1H_2H_3 g^{(4)}_{\a \a} - H_m k_\a^2$ can become negative and cause the expressions under the square roots to become negative as well. However, $R^2$ is proportional to the $\a\a$ component of the eleven-dimensional metric as $g_{\a\a}^{(11)} = R^2 (H_1 H_2 H_3)^{-2/3}$ and hence absence of CTC's outside the black hole horizon ensures that $R^2\geq 0$.}
However, unlike the point particle, this condition depends on the embedding parameters $b_1,b_2$ of the supertube through the pull-back of the metric:
\begin{equation}
g^{(4)}_{\a \a} = f (\sin^2\theta\, b_2^2 + \cos^2 \theta\, b_1^2)
+ H_m^{-1} (a_1 \cos^2\theta \,b_1 + a_2\sin^2 \theta \,b_2)^2 - (a_2 \cos^2\theta \,b_1 + a_1\sin^2 \theta \,b_2)^2\label{eq:4d_Pull-back}
\end{equation}
The ergoregion is then defined as
\begin{equation}
r_+ < r < r_{\rm erg} (\theta,b_1,b_2)\,.
\end{equation}
where $r_{\rm erg}$ is now the root of $H_m g^{(4)}_{\a \a}$. It is straightforward to see that $H_m g^{(4)}_{\a \a} >0$ when $H_m>0$; this follows immediately because the sum of the first and last terms in \eqref{eq:4d_Pull-back} is positive.
Hence the supertube ergoregion is contained in the ergoregion of point particles. Thus, it is possible for a supertube to be static when a point particle is being frame dragged!  See Figure \ref{fig:Ergoregions} for some elucidating plots.

A similar phenomenon can occur for point particles in four-dimensional background as well, when the mass of the point particle depends on the position. A position-dependent mass is generic for wrapped brane probes. One can in principle obtain such point particles by dimensional reduction of the supertube along its worldvolume, giving a point particle in a non-extremal rotating D0-D2-D6 black hole.\footnote{For dimensional reduction, one needs to consider the generalization of the Cvetic-Youm black hole to $\mathbb{R}^{1,3}\times S^1$ asymptotics first. So far, the most general non-extremal rotating black hole solution of the four-dimensional STU model has only D0-D4 charges (and charge configurations related by dualities) \cite{Chong:2004na}. Static non-extremal black holes in four-dimensions are the D0-D4 \cite{Gibbons:1982ih,Galli:2011fq}, D0-D2-D6  \cite{Galli:2012pt}, and solutions with more charges are implicitly contained in the integration algorithm of \cite{Chemissany:2009hq,Chemissany:2010zp} and the H-FGK formalism of \cite{Meessen:2011aa,Galli:2012pt}.}

\section{Phase space of supertube-black hole bound states} \label{s:PhaseSpacePlots}

In the supersymmetric limit, the background black hole becomes the BMPV solution and the configurations are supersymmetric two-center bound states. These were constructed in the probe limit first in \cite{Bena:2004wt,Marolf:2005cx}  and the fully back-reacted solution has been known for years \cite{Bena:2008dw,Bena:2011zw}. Both the probe treatment and the supergravity back-reaction show that the supertube settles at a radius set by the `bubble equation':
\begin{equation}
\frac{q_1 q_2}{d_3^2} = H_3 g_{\a\a}^{(4)} = \left(r^2 + Q_3\right) (b_1^2 \cos^2 \theta + b_2^2 \sin^2 \theta)\,,\label{eq:Bubble}
\end{equation}
where $g_{\a\a}^{(4)}$ is the pull-back of the base space metric given  \eqref{eq:4d_Pull-back} in the supersymmetric limit with $a_1\to0,a_2 \to 0$.
As the supertubes are limits of black rings with vanishing entropy, these are toy models of rings that sit at a stable distance from the black hole. Interestingly, this configuration is also related to a pure black ring by spectral flow~\cite{Bena:2011zw}.

In \cite{Bena:2005zy}, the authors showed that in the canonical ensemble, black rings and supertubes can be adiabatically brought to the horizon of a BMPV black hole by varying the transverse angular momentum of the supertube, $j_\perp$, such that the end product is again a BMPV black hole with $|J_1| = |J_2|$. This is due to a flat direction in the potential, which can extend from spatial infinity to the black hole horizon for certain charges (the bubble equation \eqref{eq:Bubble} allows a one-dimensional space of equilibrium separations). At non-zero temperature, the flat direction gets lifted and hence the question of moving a supertube into the black hole adiabatically is not well-posed. Therefore we pick charges such that the buble equation gives a flat direction that cannot extend into the black hole.
For a non-extremal black hole, the flat direction gets lifted to an isolated minimum outside the horizon. We consider the possible transition between those bound states at isolated minima through tunneling.

\begin{figure}[htb!]
\centering%
\subfigure[Supersymmetric background ($\hat m = 0$). Varying the background angular momentum changes the form of the potential, but not the position of the minimum. \label{fig:NearExtremal_Hamiltonian_a}]{
\includegraphics[width=.45\textwidth]{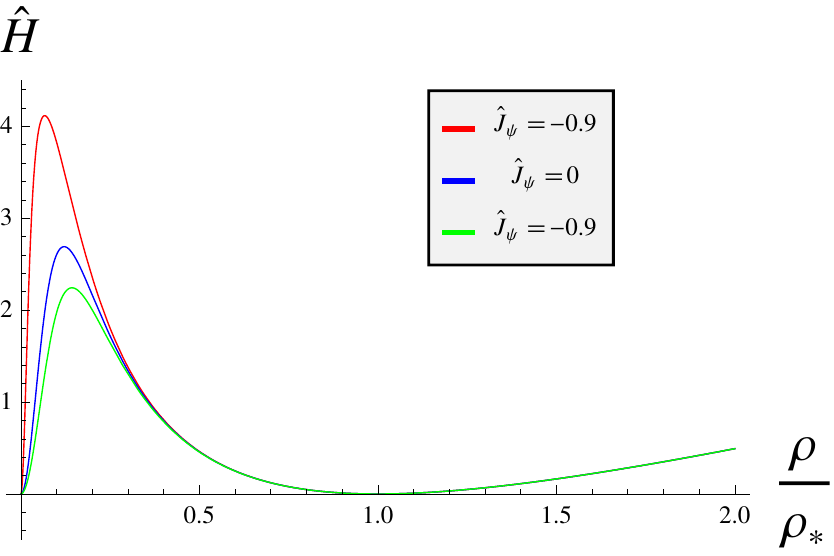}
}
\hspace{.05\textwidth}
\subfigure[Non-extremal background with $\hat m=1$. Generically, the supersymmetric minimum gets lifted. For an intermediate $\hat m$-range, local minima are possible, that can even be dynamically stable, at lower energy than the horizon value (red, solid line).\label{fig:NearExtremal_Hamiltonian_b}]{
\includegraphics[width=.45\textwidth]{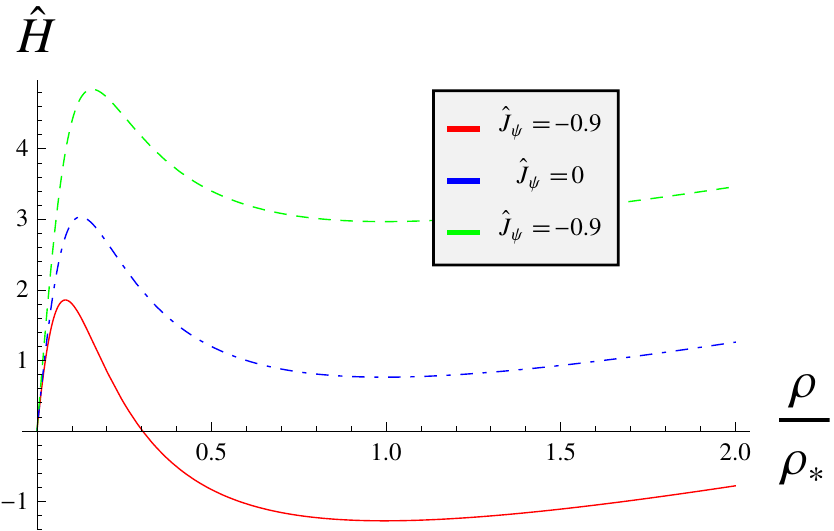}
}
\caption{The supertube Hamiltonian in the black hole background.  We use the embedding $b_1=1,b_2=0,\theta =0$ and the rescaled variables of section \ref{ss:ParameterSpace}, with all charges equal $\hat Q \equiv \hat Q_1 = \hat Q_2 =\hat Q_3,\, \hat q \equiv q_1 = q_2$. We choose $\hat d_3 = 1$ and self-dual angular momenta $\hat J_\phi = - \hat J_\psi$ of the background. We rescaled the potential $\hat H = \calh/q$ and plot versus the radial coordinate $\rho = \sqrt{r^2 - r^2_+}$. The horizon is at $\rho =0$. The supersymmetric minimum sits at the value $\rho =\rho_* \simeq 9949.87$ obtained from the bubble equation \eqref{eq:Bubble}.
\label{fig:NearExtremal_Hamiltonian}}
\end{figure}

In \cite{Chowdhury:2011qu}, two of us studied the physics of probe supertubes for non-zero temperature, in the non-extremal Cvetic-Youm background. We showed  that two-center bound states also exist when the black hole is no longer supersymmetric and has a non-zero Hawking temperature (see also \cite{Anninos:2011vn}). See Figure \ref{fig:NearExtremal_Hamiltonian} for a few plots of the supertube potential in the non-extremal Cvetic-Youm background in five dimensions. In the plots, we normalize the potential to zero at the horizon of the black hole. Remarkably, at low temperature (low $m$, near-extremal black hole), the marginally stable supersymmetric minima can become stable: the energy of the bound state is an absolute minimum, with a lower potential value than the at the black hole horizon. As we raise the temperature, stable bound states become only local minima and eventually disappear. Very far from extremality, there are no bound states, only the black hole exists.

In this section, we wish to investigate these supertube-black hole bound states in more detail. We also compare the bound states of a supertube and a non-extremal black hole, with the single-center non-extremal black hole that is formed by merging the supertube with the black hole. By merger we mean the black hole that results after tunneling of the supertube from its (meta)stable position into the black hole horizon. Hence the energy, charges and angular momenta of the merged configurations are the sum of the energies, charges and angular momenta of the background and the supertube probe, where the energy and angular momentum of the probe are evaluated at the radius at which the supertube potential reaches a local minimum.
We examine the parameter space of bound states and see in which regions in parameter space bound states exist and if they have more entropy than the merged black hole state.

We adopt the same terminology as in \cite{Chowdhury:2011qu}. We say that the bound state is {\em dynamically stable} when the potential at the local minimum is lower than that at the horizon. When the bound state has more entropy than the single center solution we will call it {\em thermodynamically stable}. Similarly, we will say the bound state is {\em dynamically metastable} when the potential at the local minima is higher than that at the horizon and finally, when the bound state has less entropy than the single center solution we will refer to it as {\em thermodynamically unstable}.

\subsection{Bound states, mergers and their entropies}

The Bekenstein-Hawking entropy of the Cvetic-Youm black hole is \cite{Cvetic:1996kv}
\begin{equation}
S_{BH} = 2\pi \sqrt{(J_+^{\rm max})^2 - J_+^2}+2\pi \sqrt{(J_-^{\rm max})^2 - J_-^2}\,.
\end{equation}
The angular momenta are:
\begin{align}
J_{\pm} &= \frac{J_\phi \pm J_\psi}2 = -\frac m2 (a_2 \pm a_1)(c_1 c_2 c_3 \mp s_1 s_2 s_3)\,,\nonumber\\
J_{\pm}^{\rm max} &=\frac{m^{3/2}}2 (c_1 c_2 c_3\mp s_1 s_2 s_3)\,.
\end{align}
A single-center black hole exists when there are no closed timelike curves outside the horizon, or equivalently when the entropy has no imaginary part. This happens when the angular momenta obey the ``cosmic censorship bounds'':
\begin{equation}
|J_\pm|\leq |J_\pm^{\rm max}| \,.
\end{equation}
In the supersymmetric limit $J_-=0$ and the bounds reduce to $|J_+| \leq \sqrt{Q_1 Q_2 Q_3}$.

To find the entropy of the bound state, we consider the Bekenstein-Hawking entropy of the background black hole only, since a supertube is a fundamental object without entropy. We will compare this to the entropy of the black hole that is formed after tunneling of the supertube into the background black hole. The charges of the merged state are
\begin{equation}
Q^{\rm tot}_{1,2} = Q_{1,2} + q_{1,2}\,, \qquad Q_{3}^{\rm tot} = Q_{3}\,,
\end{equation}
and the angular momenta are
\begin{equation}
J_\phi^{\rm tot} = J_\phi + j_\phi\,,\qquad J_\psi^{\rm tot} = J_\psi + j_\psi\,.
\end{equation}
We evaluate the supertube angular momenta $j_i$ at the local minimum of the supertube potential. Then the merger of the supertube and the black hole describes the black hole that results from tunneling of the supertube into the background black hole. Note that this is again a Cvetic-Youm black hole with only electric charges $Q_I^{\rm tot}$: since the charge $d_3$ of the supertube is a dipole charge, it does not contribute to the asymptotic charges of the black hole.

\subsection{Parameter space}

We want to understand the parameter space of black hole-supertube bound states. There are nine parameters: six for the black hole ($m,Q_1,Q_2,Q_3, J_\phi,J_\psi$) and three for the supertubes ($d_3,q_1,q_2$). Inspired by \cite{Anninos:2011vn} we make a restriction of this parameter space to visualize the different regimes.

First we restrict to the `diagonal' model, all electric charges are equal:
\begin{equation}
Q \equiv Q_1 = Q_2 = Q_3 \,,\qquad q \equiv q_1 = q_2\,.
\end{equation}
Second we use the two scaling symmetries of the system. The probe potential is invariant under the two scalings $X \to \lambda_1^{n_1} \lambda_2^{n_2} X$ of the charges $X$ (see also \cite{Anninos:2011vn}):
\begin{center}
\begin{tabular}{|l|ccc|cc|}
\hline
&$m$&$Q$&$J_i$&$d_3$&$q$\\
\hline
$n_1$&2&2&3&1&2\\
$n_2$&0&0&0&1&1\\
\hline
\end{tabular}
\end{center}
The first scaling is an invariance of the equations of motion of five-dimensional $N=2$ supergravity under conformal length rescalings. It maps a background black hole solution to another black hole solution. The second one only affects the probe charges. Both scalings affect the potential by a total conformal factor and do not change the physics.

We will use the scaling symmetries to eliminate the freedom of the charges $d_3$ and $Q$, and define scale invariant charges as $\hat X = Q^{\frac{n_2-n_1}2} {d_3}^{-n_2} X$. In particular we choose
\begin{equation}
\hat Q = 1\,, \quad \hat m = \frac m Q\,, \quad\hat J_i = \frac{J_i}{Q^{3/2}}\,,  \quad \hat d_3 = 1\,,\quad\hat  q = \frac {q} {d_3 Q^{1/2}}\,.
\end{equation}
This leaves us with a four-dimensional parameter space. We make two-dimensional slices of phase space by additionally fixing the ratio $J_\phi/J_\psi$ and  the probe charge.

Note that the probe approximation is valid when the probe mass is small compared to the background mass: $m_p \ll M$. Since the ratio of these two masses has the same scaling behaviour as the ratio of the probe and background electric charges, we have:
\begin{equation}
\frac{m_p}M = \frac{d_3}{Q^{1/2}} \frac{\hat m_p}{\hat M}\,.\label{eq:ProbeMassRatio}
\end{equation}
By making the ratio $d_3/Q^{1/2}$ small, we can always make sure the probe regime is valid.

\subsection{Scans of parameter space}\label{ss:ParameterSpace}

To study the existence of metastable and stable bound states, we perform a numerical scan of parameter space. We choose the probe charges and charge ratio
\begin{equation}
\hat q = 10\,,
\end{equation}
and the probe embedding
\begin{equation}
b_1 = 1\,,\qquad \ b_2 = 0\,,\qquad \theta = 0\,.
\end{equation}
With this choice of embedding, the local minima of the potential are at $\sin \theta = 0$ due to symmetry.

The total charges are
\begin{equation}
 Q^{\rm tot}_{1,2} =  Q +  q\,, \qquad  Q_{3}^{\rm tot} =  Q\,,
\end{equation}
and the angular momenta are (see \eqref{eq:AngMom_Parallel} and \eqref{eq:AngMom_perp_1})
\begin{equation}
\hat J_\phi^{\rm tot} = \hat J_\phi + \frac{d_3}{ Q^{1/2}}\,,\qquad \hat J_\psi^{\rm tot} = \hat J_\psi + \frac{d_3}{ Q^{1/2}}\hat q^2\,.
\end{equation}
The remaining parameter space is four-dimensional: the three rescaled variables $(\hat m, \hat J_\phi, \hat J_\psi)$ and the charge ratio $d_3/Q^{1/2}$ that fixes the probe-to-background mass ratio (see \eqref{eq:ProbeMassRatio}). For illustrative purposed, we only make plots of phase space for one value of this ratio. We fix:
\begin{equation}
d_3/Q^{1/2} = 10^{-3}\,.
\end{equation}
Other values do not change the qualitative observations. We perform two 2-dimensional slicings, one with self-dual angular momenta $J_\phi = -J_\psi$, one with $J_\phi = 0$. The self-dual angular momenta have a well-defined supersymmetric limit $\hat m \to 0$ keeping the charges at fixed positive values.

\subsubsection{Background with self-dual angular momenta}

We first consider a background with  self-dual angular momenta:
\begin{equation}
\hat J_\phi = -\hat J_\psi\,.
\end{equation}

\begin{figure}[htp!]
\centering
\subfigure[Background angular momenta $J_\psi = - J_\phi$\label{fig:Scan_J_selfdual}.]{
\includegraphics[width=.49\textwidth]{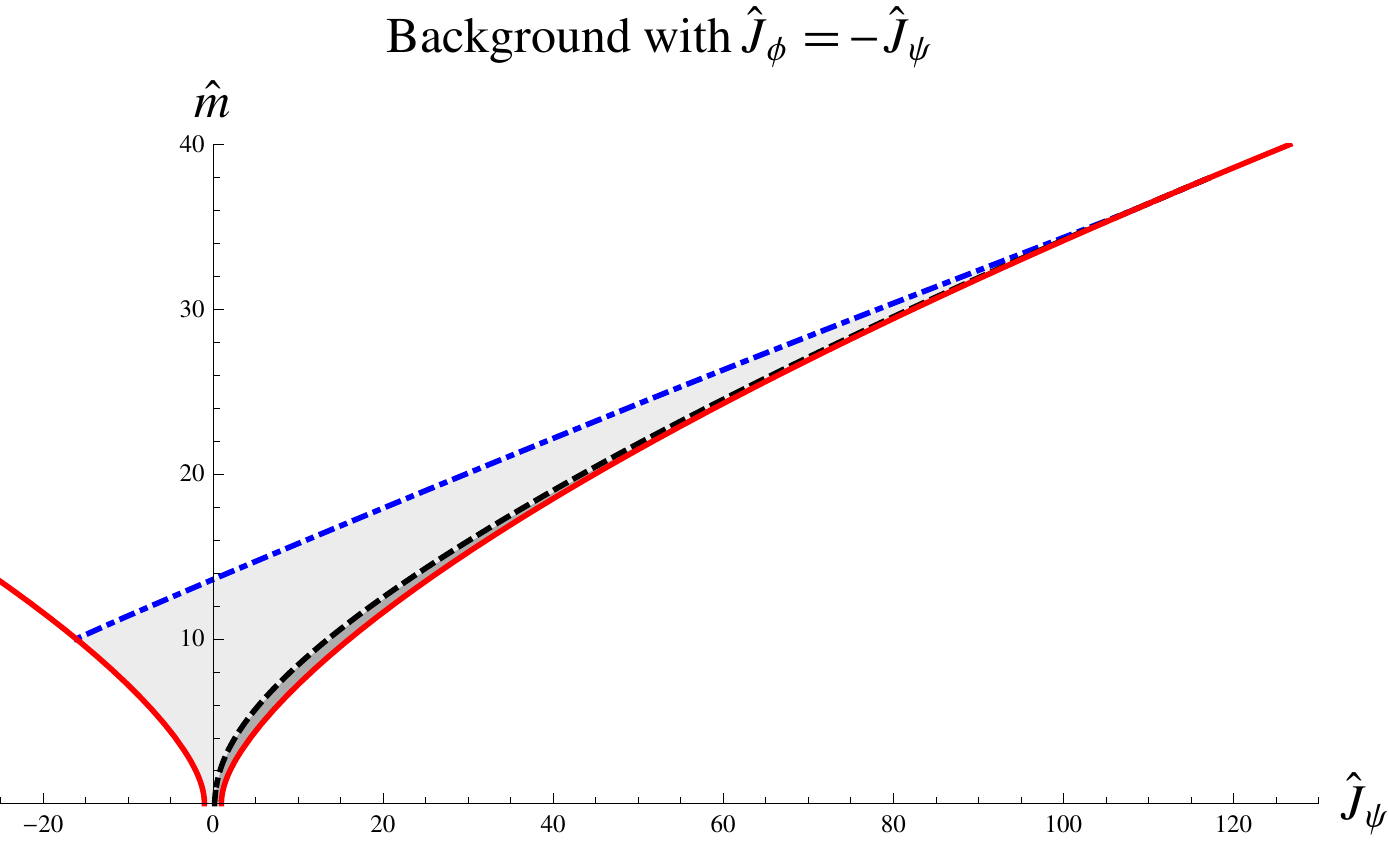}
\includegraphics[width=.49\textwidth]{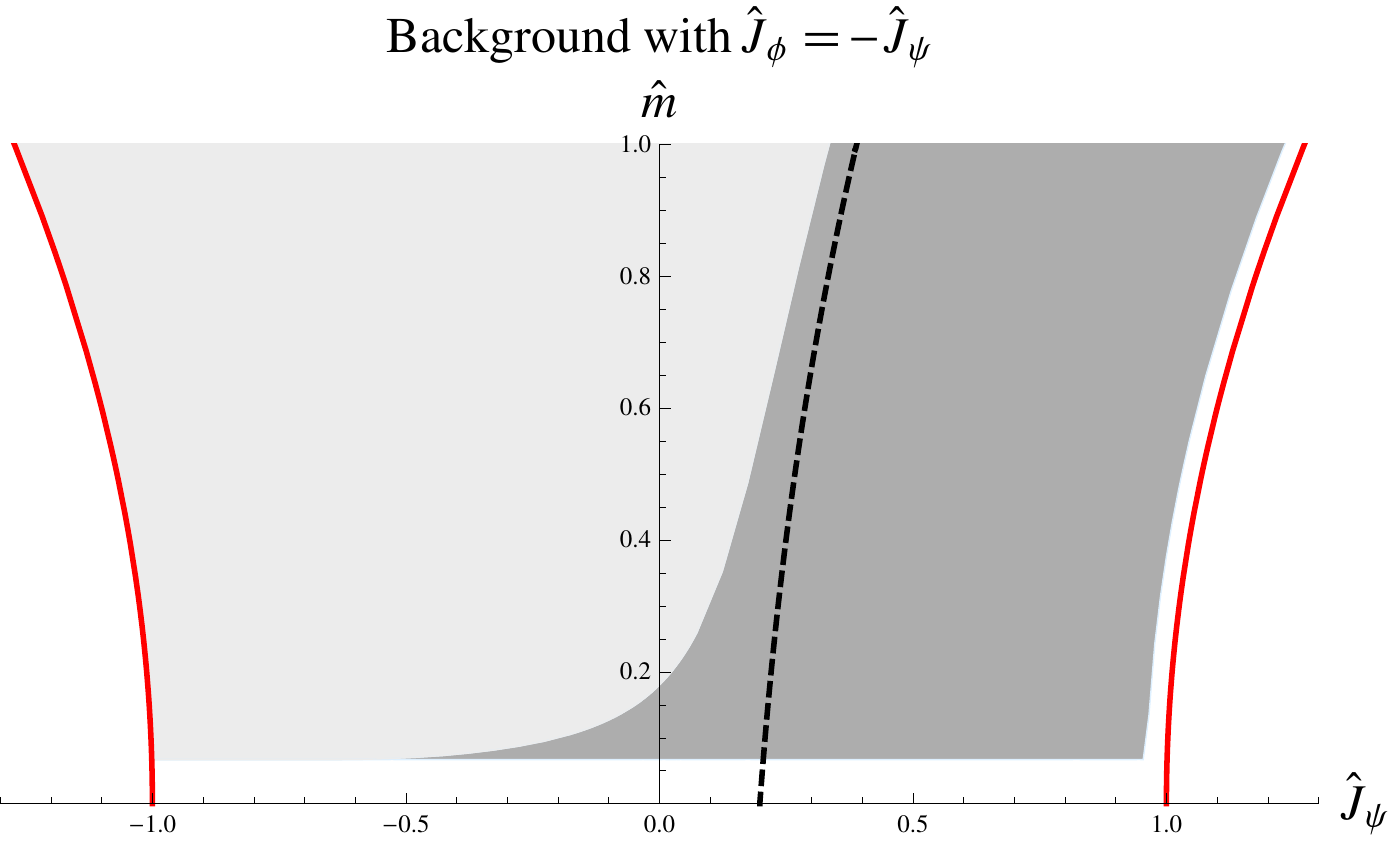}}
\subfigure[Background with one angular momentum ($\hat J_\phi =0$)\label{fig:Scan_J_zero}]{
\includegraphics[width=.49\textwidth]{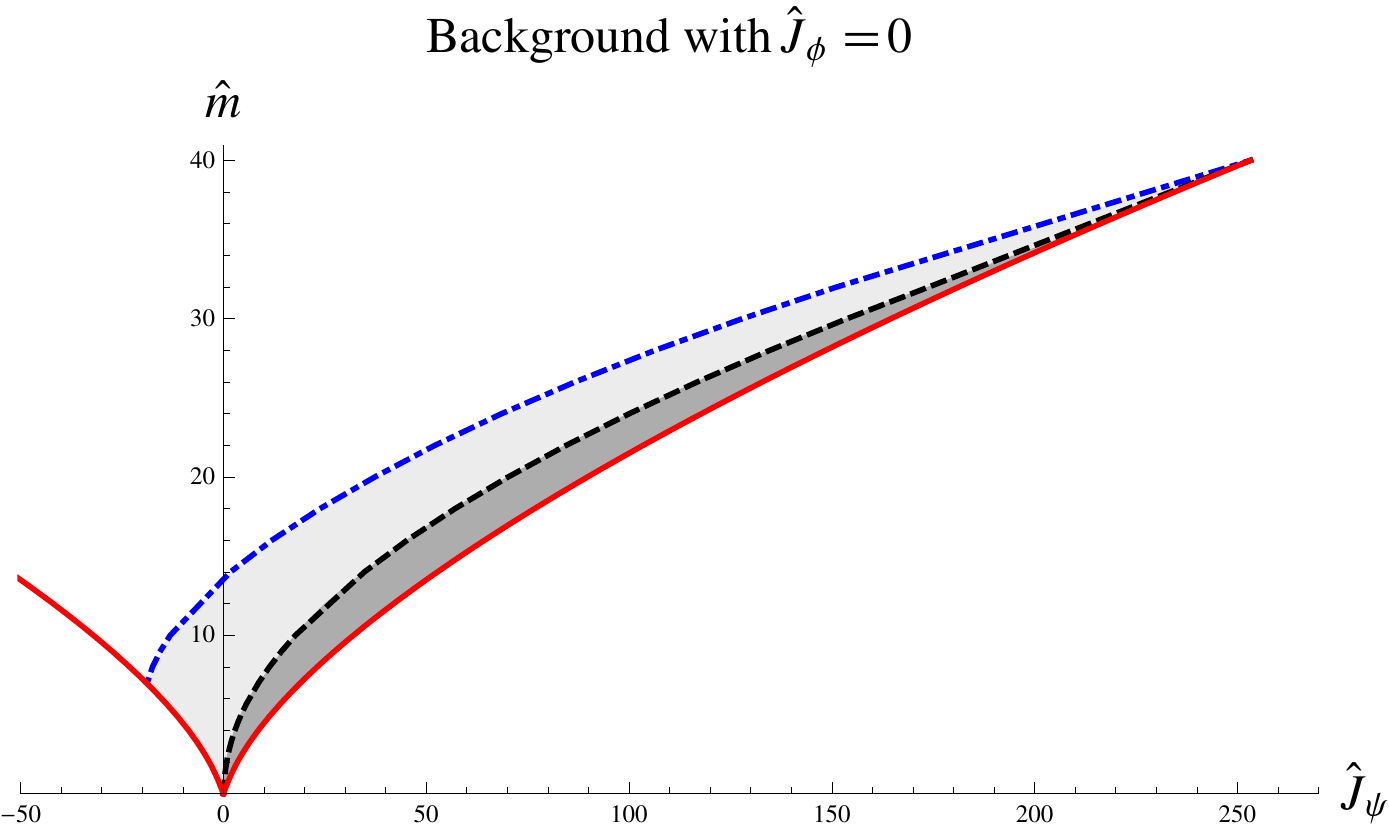}
\includegraphics[width=.49\textwidth]{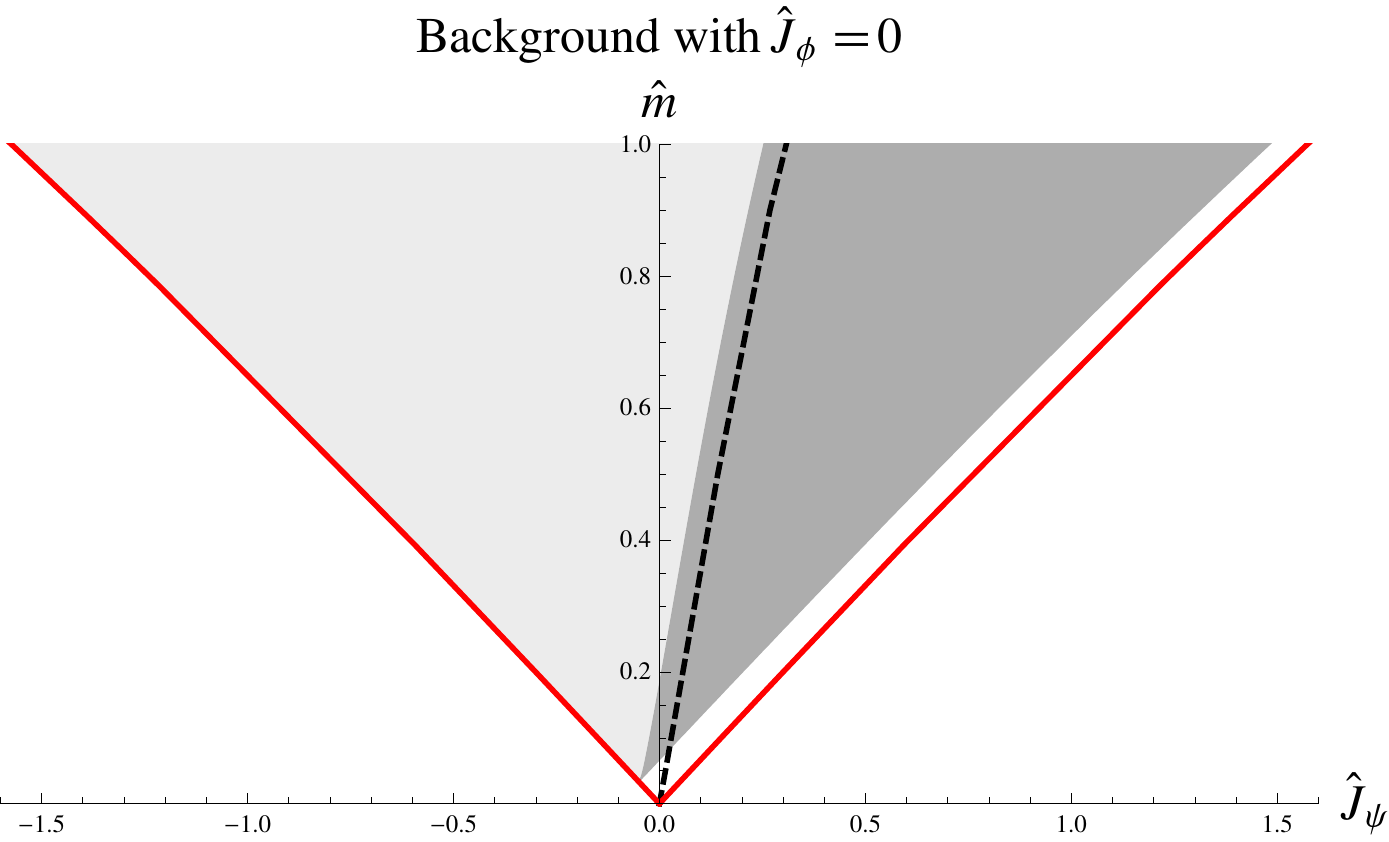}
}
\subfigure[Legend]{
\includegraphics[width=.55\textwidth]{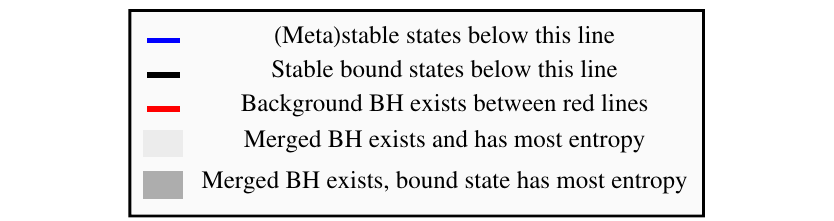}
}
\caption{Figure \protect\subref{fig:Scan_J_selfdual}: Self-dual angular momenta of the background. The right graph is a zoom of the bottom region of the left one. The black hole background exists for given $m$ for angular momenta in between the red (solid) lines. Bound states exist between the blue (dash-dot) and the red (solid) lines. Dynamically stable bound states exist between the black (dashed) and the red (solid) lines. In the dark-grey region, the bound state has more entropy than the merger of the background black hole with the supertube (i.e. the bound state is thermodynamically stable). In the light-grey region, the merger is most entropic.
Figure \subref{fig:Scan_J_zero}: The background black hole has $\hat J_\phi = 0$. The right graph is a zoom of the bottom region of the left one. Bound states exist between the blue (dash-dot) and the red (solid) lines. Dynamically stable bound states exist between the black (dashed) and the red (solid) lines. In the dark-grey region, the bound state has more entropy than the merger of the background black hole with the supertube (i.e. the bound state is thermodynamically stable). In the light-grey region, the merger is most entropic.
\label{fig:scan_J}}
\end{figure}

We examine the phase space of supertube bound states in the $(\hat J_{\psi},\hat m)$-plane of the background black hole, in Figure \ref{fig:Scan_J_selfdual}. Note that the line $\hat m =0$ for $|\hat J_\psi | \le 1$ corresponds to BMPV black holes.

%---------FIGURE------------
%---------End-FIG-----------

We see that from the global picture we may conclude that \emph{thermodynamic stability goes hand in hand with dynamical stability}. The boundary between the thermodynamically stable and metastable states (boundary between light-grey and dark-grey regions) follows closely the boundary between the regions in phase space with dynamically stable and dynamically metastable bound states (black, dashed line). All dynamically stable bound states are also thermodynamically stable compared to the black hole with the same total charges that describes the merger of the background with the supertube. On the other hand, most dynamically metastable states are thermodynamically unstable compared to the merged black hole.

We observe two very interesting exceptions to the general observation. The first is that right to the left of/above the black, dashed line, there is a thin dark-grey band of dynamically metastable states that are nevertheless thermodynamically stable over the merged black hole. Even though the two-center potential tells us that the supertube at the horizon has lowest energy, the larger entropy of the bound state compared to the merged state shows that the supertube and the black hole do not form a stable single-center end product for the charge configurations in this small band. It is likely that the correct end point for this set of charges is some other multi-centered configuration. However, the end point is definitely not a single center black hole.

For the second interesting exception, note that there is a small white band near the cosmic censorship bound of the background black hole (red, solid line) for positive $\hat J_{\psi}$ and for low background temperature (low $\hat m$), where the merger itself is forbidden by cosmic censorship. For dynamically metastable bound states (on the left of the black, dashed line) this result is surprising because one would expect everything to fall into a black hole. Nevertheless, the black hole cannot absorb the supertube on account of shortage of phase space. We expect that for dynamically metastable bound state in this region, there are other decays products of black holes and supertubes with other charge channels, or possibly more supertubes. For dynamically stable bound states in this region (on the right of the black, dashed line) there is no reason to expect any other phase. However, it is interesting to note that we cannot think of the supertube being ``spat out" in this case as there is no corresponding single center black hole.

\subsubsection{Background with rotation in one plane}

We consider a background where one of the angular momenta of the background black hole is zero
\begin{equation}
J_\phi = 0\,.
\end{equation}
The non-zero angular momentum is then along the cycle on which the supertube is wrapped. We plot the phase space in Figure \ref{fig:Scan_J_zero}). The findings are qualitatively the same as for a background with self-dual angular momenta. The quantitative differences are that thermodynamically stable bound states exist up to $\hat J_\psi =0$ as the background temperature goes to zero\footnote{At first sight, it might seem confusing that e.g. the black, dashed boundary line (separating dynamically stable and dynamically meta-stable states) ends in the point $\hat{J}_{\psi}=0$ for $\hat{m}\rightarrow 0$ in the graph where $\hat{J}_{\phi}=0$, while for $\hat{J}_{\phi}=-\hat{J}_{\psi}$, the intersection with $\hat{m}=0$ seems to be at finite $\hat{J}_{\phi}$. However, strictly speaking, there are no dynamically bound states for $\hat{m}=0$ so that the black, dashed line is discontinuous at the point $\hat{m}=0$ for the graph $\hat{J}_{\phi}=-\hat{J}_{\psi}$; such dynamically bound states only appear as soon as we add even the smallest bit of self-dual angular momentum.}, and that the small band of dynamically metastable bounds states that are thermodynamically stable compared to the merger does not significantly widen at low $\hat m$.

\section{Conclusions} \label{s:Conclusions}

Multi-centered black hole bound states exhibit rich physics and have been important in shedding light on various aspects of supergravity and string theory. There has been progress in the construction of multi-centered bound states in the case of non-extremal configurations recently. On the probe level, one can establish that supertubes form locally stable bound states with the non-extremal black hole~\cite{Anninos:2011vn,Bena:2011fc,Chowdhury:2011qu,Bena:2012zi,Chowdhury:2013pqa}, which serve as testing grounds for more intricate bound states of black holes and black rings. While the aforementioned papers studied the dynamics for a probe center around a non-extremal black hole, in this paper we went beyond dynamics to study the thermodynamics of such multi-centered solutions.

Our work  confirms earlier conjectures based on the D1-D5 decoupling limit of \cite{Chowdhury:2011qu}, where we compared stability and metastability of probe branes to entropies of the dual CFT states at the orbifold point. In this paper we can study entropies directly in the gravitational description that is dual to a strongly coupled CFT unlike the weakly coupled CFT at the orbifold point. We find that  dynamical stability (potential at the local minimum of the supertube potential lower than that at the horizon) implies thermodynamic stability (entropic dominance of the bound state over the black hole with the same total charges). In the other direction, the connection is less strong. Dynamic metastability (potential at the local minimum of the supertube potential higher than that at the horizon) is largely synonymous with thermodynamic instability (entropic dominance of the merged state over the bound state with the same total charges). However, there is band of dynamically metastable states which are nevertheless thermodynamically stable. In addition there is another band of states, comprising both dynamically metastable and dynamically stable states, which have no corresponding merged single center black hole.

These exceptions lead us to believe that there are some dynamically metastable configurations which could tunnel into other multi-centered configurations but not single center black holes. This indicates a very rich physics for non-extremal multi-centered solutions.

Through T-dualities and spectral flow, the supertube-black hole bound states we have studied in this paper should be dual to black rings in Taub-NUT, where the charge of the ring is large compared to the Taub-NUT charge (see \cite{Bena:2011zw} for this transformation in a supersymmetric setup). Hence we can `predict' the existence of new non-extremal black rings in Taub-NUT space! The curious fact is that the Taub-NUT center becomes a probe in our setup. However, by playing with the ratio of probe and background charges, we can have integer charges of the probe and still be well in the probe regime.

It would be very interesting to further explore the phase space of charged multicenter bound states, similar to the phase structure of five-dimensional black holes and black rings in GR \cite{Elvang:2007hg,Emparan:2007wm}. One straightforward application is to study bound states in minimal supergravity in five dimensions (three equal M2 charges from the eleven-dimensional point of view). This is the charge setup of the four-dimensional probes used in \cite{Anninos:2011vn}, the four-dimensional $t^3$-model. This theory has a restricted set of parameters such that it becomes possible to study phase diagrams explicitly, but it is still rich in physics. In particular, the probes in this theory are no longer supertubes. Note that there is a possible complication, since in principle the DBI action for probes in thermal backgrounds needs to be corrected as in \cite{Grignani:2010xm,Grignani:2011mr,Grignani:2012iw,Armas:2012bk,Armas:2013ota}.

Methods such as the blackfold approach \cite{Emparan:2009cs,Emparan:2009at} can complement our probe approximation. As we noted above, the back-reaction of the supertube-black hole bound states of our current analysis can be related through T-dualities and spectral flow to a very massive non-extremal black ring with tree electric and three dipoles charges in Taub-NUT. Such black rings can be treated as blackfolds in a certain regime, depending on the ratio of the size of the Taub-NUT circle and the thickness of the ring horizon. To treat with such solutions of five-dimensional supergravity, the blackfold approach needs to be extended first to theories with Chern-Simons couplings of the gauge fields.

\section*{Acknowledgments}

We would like to thank J.\ Armas, I.\ Bena, J.\ de Boer, F.\ Denef, S.\ El-Showk, F.\ Larsen, N.\ Obers, M.\ Shigemori and T.\ Van Riet for stimulating discussions and especially I.\ Bena and J.\ de Boer for useful comments on the draft. We also thank the referee for constructive comments. BV is supported by the ERC Starting Independent Researcher Grant 240210 - String-QCD-BH and would like to thank the Pedro Parcual Centro de Ciencias in  Benasque, the organizers of the Benasque workshop on ``Gravity: perspectives from strings and higher dimensions'' and its participants for discussions and suggestions. This work is part of the research programme of the Foundation for Fundamental Research on Matter (FOM), which is part of the Netherlands Organisation for Scientific Research (NWO).

\appendix{

\section{Gauge parameters for a probe string}\label{app:SO(4)}

We write the Euclidean coordinates of four-dimensional flat space as
\begin{eqnarray}
x_1 = \sin \theta \cos \phi\,,&\qquad & x_3 = \cos \theta \cos \psi\,,\nonumber\\
x_2 = \sin \theta \sin \phi\,,&\qquad & x_4 = \cos \theta \sin \psi\,,
\end{eqnarray}
We get the following rotation generators $\xi_{ij} = x_i\partial_j - x_j\partial_i$:
\begin{eqnarray}
\xi_{12} &=& \partial_\phi\,,\nonumber\\
\xi_{34} &=& \partial_\psi\,,\nonumber\\
\xi_{23} &=& - \sin \phi \cos \psi\, \partial_\theta -  \cot \theta \cos \phi \cos \psi\,\partial_\phi - \tan \theta \sin \phi \sin \psi \,\partial_\psi\,,\nonumber\\
\xi_{13} &=& - \cos \phi \cos \psi\, \partial_\theta +  \cot \theta \sin \phi \cos \psi\,\partial_\phi - \tan \theta \cos \phi \sin \psi \,\partial_\psi\,,\nonumber\\
\xi_{14} &=& -\cos \phi \sin \psi\, \partial_\theta + \cot \theta \sin \phi \sin \psi\, \partial_\phi + \tan \theta \cos \phi \cos \psi\,\partial_\psi\,,\nonumber\\
\xi_{24} &=& - \sin \phi \sin \psi \,\partial_\theta - \cot \theta \cos \phi \sin \psi\, \partial_\phi + \tan\theta \sin\phi \cos \psi\,\partial_\psi \,.
\end{eqnarray}
They satisfy the $SO(4)$ algebra:
\begin{equation}
[\xi_{ik},\xi_{j\ell}] = \delta_{i\ell}\xi_{kj}- \delta_{k\ell}\xi_{ij}- \delta_{kj}\xi_{i\ell}- \delta_{ij}\xi_{k\ell}\,.
\end{equation}

The Lie derivative of $C_2=m(k-\cos^2 \theta)$ determines the exterior derivative of the gauge one-forms $\Lambda$ through \eqref{eq:GaugeTransfo_PForm}:
\begin{equation}
\call_{\xi_{ij}} C_2 \equiv d\Lambda^{ij}\
\end{equation}
A set of one-forms that satisfies this condition is:
\begin{eqnarray}
\Lambda^{12} &=& -k \,d\psi \,, \nonumber\\
\Lambda^{13} &=& m \sin\phi \sin\psi \,d\theta -(k+m) \cos\psi \cot\theta \sin\phi \,d\psi - k \cos\phi \sin\psi \tan\theta \,d\phi\,, \nonumber\\
\Lambda^{14} &=& -m \cos\psi\sin\phi \,d\theta-(k + m)\cot\theta\sin\psi\sin\phi\,d\psi+ k \cos\phi \cos\psi \tan\theta \,d\phi\,, \nonumber\\
\Lambda^{23} &=& -m \cos\phi \sin\psi \,d\theta +(k+m) \cos\phi \cos\psi \cot\theta \,d\psi- k \sin\phi \sin\psi \tan\theta \,d\phi\,, \nonumber\\
\Lambda^{24} &=& m\cos\phi\cos\psi\,d\theta+(k + m)\cot\theta\sin\psi\cos\phi \,\,d\psi  +k \cos\psi \sin\phi \tan\theta \,d\phi \,, \nonumber\\
\Lambda^{34} &=& (k+m) \,d\phi\,.
\end{eqnarray}
The arbitrary choice for $\Lambda^{12},\Lambda^{34}$ was fixed by demanding that these gauge one-forms obey the condition \eqref{eq:Lambda_Gauge_orig}:
\begin{equation}
i_{\xi_B} d\Lambda_A-i_{\xi_A} d\Lambda_B =  f_{AB}{}^C(\Lambda_C + d \lambda'_C)\,.
\end{equation}
With the current choice of $\Lambda^I$, the gauge transformations $\lambda_C$ are
\begin{equation}
f_{AB}{}^C \lambda_C=-\frac 12 (\xi_A^i \Lambda_i^B - \xi^i_B \Lambda_A^i)\,.
\end{equation}
The conserved charges $Q^{ij} = \xi_{ij}^kp_k + \Lambda^{ij}$ satisfy the Poisson brackets for $\SO(4)$.

\section{The Probe Hamiltonian and angular momentum}\label{app:AngularMomentum}
The procedure to find the probe Hamiltonian is very similar to that described in appendix B of \cite{Chowdhury:2011qu}. We will only sketch the procedure here and highlight the differences with \cite{Chowdhury:2011qu}.

\subsection{Probe Lagrangian}
We write the background metric as
\begin{eqnarray}
ds_{11}^2 &=& -(H_1 H_2 H_3)^{-2/3} H_m (dt + k)^2 + (H_1 H_2 H_3)^{1/3} ds_4^2 + \sum_{I=1}^3 \frac{(H_1 H_2 H_3)^{1/3}} {H_I}ds_I^2,
\end{eqnarray}
and we introduce coordinates on the three torii as $(z,x_{11})$, $(y_1,y_2)$ and $(y_3,y_4)$:
\begin{equation}
ds_1^2 = dz^2 + dx_{11}^2\,,\quad ds_2^2 = dy_1^2 + dy_2^2\,,\quad ds_3^2 = dy_3^2 + dy_4^2\,.
\end{equation}
The probe is a supertube, consisting of an M5-brane with dissolved M2-branes. The M5-brane wraps the coordinates of the first two $T^2$'s ($x_{11},z,y_1,y_2$), as well as a direction in the non-compact space. Two M2-branes are dissolved in the M5, they are wrapped on torus 1 and torus 2. To find the Hamiltonian description of this M5-brane, it is easiest to first reduce to 10D type IIA supergravity on the direction $x^{11}$. The M5-brane probe becomes a D4-brane, for which the action is:
\begin{align}
S_{D4} &= S_{DBI} + S_{WZ},\\
S_{DBI} &= -|N_{D4}| T_{D4}\int d^5\xi e^{-\Phi}\sqrt{-\det\left(g+B+F\right)},\\
S_{WZ} &= N_{D4} T_{D4} \int d^5\xi \left( C_5 + C_3\wedge(B+F)\right).
\end{align}
The embedding is given by $\xi^0\equiv\tau=t,\xi^1=z,\xi^2=\alpha,\xi^3=y_1,\xi^4=y_2$ and:\footnote{We use the world-volume Levi-Civita symbol convention $\epsilon^{\xi^0\xi^1\xi^2\xi^3\xi^4}=+1$.}
\begin{equation}
\psi = b_1 \a + v_1\tau\,, \quad\phi = b_2 \a+v_2\tau\,.
\end{equation}
The parameters $v_i$ (which are new with respect to the discussion in \cite{Chowdhury:2011qu}) determine the angular velocity of the supertube. We will set these to zero in the end since we are interested in static supertubes. They are needed to determine the angular momenta of the tube, as we will see shortly.

The metric, dilaton, NS-NS form $B_2$, and R-R form $C_3$ in 10D can be read off easily from the 11D background \ref{eq:11d_Background} (or from \cite{Chowdhury:2011qu}). The relevant components of $C_5$ can be obtained by dualizing $C_3$ using $dC_5 = -*dC_3-H_3\wedge C_3$ (since $C_1=0$), and are given by:\footnote{We use the 10D convention  $\epsilon_{t r \theta \phi\psi z y_1 y_2 y_3 y_4} = +1$ for the Levi-Civita symbol.}
\begin{eqnarray}
C_{t \psi z12} &=&  \f{m \cos^2 \theta}{f H_1} (a_2 c_1 c_2 s_3 - a_1 s_1 s_2 c_3)\,, \\
C_{t \phi z12} &=&  \f{m \sin^2 \theta}{f H_1} (a_1 c_1 c_2 s_3 - a_2 s_1 s_2 c_3)\,,\\
C_{\psi\phi z12} &=& - \f{Q_3}{f H_2}\left[ (r^2 + a_2^2 + m s_2^2) \cos^2 \theta - \f{m(s_2^2-s_1^2)}{fH_1}(a_1^2-a_2^2)\cos^2 \theta \sin^2 \theta\right]\,.
\end{eqnarray}
Finally, the world-volume field on the D4-brane is given by:
\begin{equation} F = \mathcal{E}\, d\xi^0\wedge d\xi^1 + \mathcal{B}\, d\xi^1\wedge d\alpha.\end{equation}
The electric field $\mathcal{E}$ is a source for F1 charge in the D4 worldvolume while the magnetic field $\mathcal{B}$ is a source for D2 charge.

After some algebra, one finds the Born-Infeld and Wess-Zumino Lagrangians are:
\begin{eqnarray}
\call_{BI} &=& - T_{D4}  \frac{Z}{H_2}\Bigg{[} \frac{Z^3}{H_1^2}\left(Z((g_{\a\tau}^{(4)})^2 - g_{\a\a}^{(4)}g_{\tau\tau}^{(4)}) + \frac{H_m}{Z^2}[g_{\a\a}^{(4)} (1+k_\tau)^2 + k_\a(g_{\tau\tau}^{(4)} k_\a - 2 g_{\a\tau}^{(4)}(1+k_\tau))]\right)\nonumber\\
&&\quad \qquad+\frac{H_m}{Z^2}[(1+k_\tau)\tilde \calb + k_\a \tilde \cale]^2 - Z(\tilde \cale^2 g_{\a\a}^{(4)}+ 2 \tilde \calb \tilde \cale g_{\a\tau}^{(4)} + \tilde \calb^2 g_{\tau\tau}^{(4)})\Bigg{]}^{1/2}\,,\\
\call_{WZ} &=& T_{D4}(C_{t \a z 3 4} + (v_1 b_2 - v_2 b_1)C_{\psi\phi z 34} - \tilde \calb C_{\tau 34} - \tilde \cale C_{\a 34})\,,
\end{eqnarray}
where we remind the reader that $g^{(4)}$ is the four-dimensional base metric \eqref{eq:4d_Base} and $k$ the rotation one-form \eqref{eq:kvector}.
The shifted electric and magnetic fields appearing in this expression are defined as
\begin{equation}
\tilde \cale  = (B + F)_{\tau z}\,, \qquad\tilde \calb = (B + F)_{z\a}\,,
\end{equation}
and the worldvolume components of the two-form and three-form fields are
\begin{equation}
\begin{array}{ll}
C_{\tau 34} = A_t^{(2)}+ v_1 A_{\psi}^{(2)} + v_2 A_{\phi}^{(2)}\,,\qquad &B_{\tau z} = A_t^{(1)}+ v_1 A_{\psi}^{(1)} + v_2 A_{\phi}^{(1)}\,,\nonumber\\
C_{\a34} = A_\a^{(2)} = b_1 A_{\psi}^{(2)} + b_2 A_{\phi}^{(2)}\,,&B_{z\a} = -A_\a^{(1)} =-( b_1 A_{\psi}^{(1)} + b_2 A_{\phi}^{(1)})\,.\nonumber\\
\end{array}
\end{equation}
For later use, we give the electric field at zero velocity ($v_i=0$):
\bea
\tilde \cale &=& \tilde q_2\frac{k_\a H_m}{\rho^2}  + \tilde q_1 \frac{\sqrt{H_m Z^3 g_{\a\a}}}{\rho^2} \sqrt{\frac{\tilde q_1^2 + \frac{\rho^2}{H_2^2}}{\tilde q_2^2 + \frac{\rho^2}{H_1^2}}}\label{eq:Charges_ElecField}\,,
\eea
with the shifted charges $\tilde q_1,\tilde q_2$ defined in eq.\ \eqref{ShiftedCharges}.

\subsection{Probe Hamiltonian}
The electric field $\mathcal{E}$ is not a conserved quantity, so we need to Legendre transform the Lagrangian with respect to $\mathcal{E}$ to obtain the Hamiltonian of the tube, which will depend on the conserved charges $q_1,q_2$.

The conserved F1 Page charge is given by:
\begin{equation} q_1 = \frac{\partial \mathcal{L}}{\partial \mathcal{E}}.\end{equation}
The Hamiltonian $\mathcal{H}$ is then given by:
\begin{equation} \mathcal{H} = q_1 \mathcal{E} - \mathcal{L}.\end{equation}
We further denote the D2-charge by $q_2$ and D4-dipole charge by $d_3$, so:
\begin{equation} q_2\equiv d_3 \mathcal{B}, \qquad d_3\equiv N_{D4}T_{D4}\,.\end{equation}

Then, working in units where the masses of the three tori are equal to 1 (see appendix A and B of \cite{Chowdhury:2011qu} for more details), and setting the angular velocity parameters $v_1=v_2=0$, we obtain the Hamiltonian given in (\ref{eq:NonExtremalHamFullBulk}).

\subsection{Probe angular momentum}
The background breaks rotational invariance, so there will not be a full $SO(4)$ algebra of conserved angular momenta for the supertube. However, the background (\ref{eq:11d_Background}) still has $SO(2)\times SO(2)$ symmetry generated by Killing vectors $\partial_{\phi}$ and $\partial_{\psi}$, so the angular momenta $j_{12}=j_{\phi}$ and $j_{34}=j_{\psi}$ will be conserved quantities. The angular momentum along the $x^3-x^4$ plane is given by:
\begin{equation} j_{\psi} = \frac{\partial\mathcal{L}}{\partial(\partial_{\tau}\psi)} + T_{D4} d_3 b_2 \kappa_1 = \frac{\partial\mathcal{L}}{\partial v_1}+ T_{D4} d_3 b_2 \kappa_\psi,\end{equation}
where we have added an a priori arbitrary constant to the quantity needed to fix the gauge ambiguity as discussed in section \ref{s:Gauge}. In an analogous fashion, we have:
\begin{equation} j_{\phi} = \frac{\partial\mathcal{L}}{\partial v_2}+ T_{D4} d_3 b_1 \kappa_\phi.\end{equation}
Again, after taking the partial derivatives, we set $v_1=v_2=0$; the result is the expression (\ref{eq:AngularMomentum}).

At spatial infinity, rotational invariance is asymptotically realized; so all of the angular momenta of the supertube should asymptotically be conserved and satisfy the full $SO(4)$ algebra. Equivalently, we can consider the flat space limit of the background; in this limit, we again have rotational invariance and a full $SO(4)$ algebra of conserved angular momenta for the tube. For these limits, we can thus apply the reasoning of section \ref{s:Gauge} and determine the constants $\kappa_i$ from demanding that $j_i$ are the correct generators in the $SO(4)$ algebra of conserved angular momenta; this determines them to be given as in (\ref{eq:angmomkappas}):
\be
\kappa_\psi = -Q_3\,,\qquad \kappa_\phi = 0\,.
\end{equation}

}

\providecommand{\href}[2]{#2}\begingroup\raggedright\endgroup

\end{document}